\documentclass[preprint,11pt,sort,compress]{elsarticle}

\usepackage{lmodern,sansmath}
\usepackage{textcomp}
\usepackage{microtype}
\usepackage{amsmath}
\usepackage{amssymb}
\usepackage{booktabs}
\usepackage{graphicx}
\usepackage{xcolor}

\def\dd#1#2{\frac{\mbox{d} #1}{\mbox{d} #2}}
\def\pp#1#2{\frac{\partial #1}{\partial #2}}
\renewcommand{\vec}[1]{\boldsymbol{#1}}

\newcommand{\avg}[1]{\langle{#1}\rangle}

\journal{Computers \& Fluids}

\begin{document}

\begin{frontmatter}



  \title{\textbf{On the self-similarity of line segments in decaying
      homogeneous isotropic turbulence}}


  \author[ad1]{Michael Gauding\corref{cor1}}
  \ead{michael.gauding@coria.fr}
  \cortext[cor1]{Corresponding author}

  \author[ad2]{Lipo Wang}
  \author[ad3]{Jens~Henrik~Goebbert}
  \author[ad4]{Mathis~Bode}
  \author[ad1]{Luminita~Danaila}
  \author[ad1]{Emilien~Varea}

\address[ad1]{CORIA -- CNRS UMR 6614, Saint Etienne du
  Rouvray, France}

\address[ad2]{UM-SJTU Joint Institute, Shanghai JiaoTong University,
  Shanghai, China}

\address[ad3]{Juelich Supercomputing Center, Juelich, Germany}

\address[ad4]{Institute for Combustion Technology, RWTH Aachen
  University, Germany}

\begin{abstract}
  The self-similarity of a passive scalar in homogeneous isotropic
  decaying turbulence is investigated by the method of line segments
  (M. Gauding et al., Physics of Fluids 27.9 (2015): 095102).  The
  analysis is based on a highly resolved direct numerical simulation
  of decaying turbulence. The method of line segments is used to
  perform a decomposition of the scalar field into smaller sub-units
  based on the extremal points of the scalar along a straight
  line. These sub-units (the so-called line segments) are
  parameterized by their length $\ell$ and the difference $\Delta\phi$
  of the scalar field between the ending points.  Line
    segments can be understood as thin local convective-diffusive
    structures in which diffusive processes are enhanced by
    compressive strain.  From DNS, it is shown that the marginal
  distribution function of the length~$\ell$ assumes complete
  self-similarity when re-scaled by the mean length $\ell_m$. The
  joint statistics of $\Delta\phi$ and $\ell$, from which the local
  gradient $g=\Delta\phi/\ell$ can be defined, play an important role
  in understanding the turbulence mixing and flow structure.  Large
  values of $g$ occur at a small but finite length scale. Statistics
  of $g$ are characterized by rare but strong deviations that exceed
  the standard deviation by more than one order of magnitude.  It is
  shown that these events break complete self-similarity of line
  segments, which confirms the standard paradigm of turbulence that
  intense events (which are known as internal intermittency) are not
  self-similar.
\end{abstract}





\end{frontmatter}

\section{Introduction}

The turbulent motion of fluids is a highly complex phenomenon. In
general, turbulent flows are characterized by random spatio-temporal
fluctuations over a broad range of scales.  Reliable predictions of
the statistical properties of these fluctuations will be of practical
importance for a wide field of applications.

Historically, turbulence research has mostly focused on a statistical
description in the sense of Kolmogorov's scaling theory
\cite{kolmogorov1941b,kolmogorov1941}, which hypothesizes that for
sufficiently large Reynolds numbers, the small scale motion is
statistically independent from the large scales. While the large
scales depend on the boundary or initial conditions, the smallest
scales should be statistically universal and feature certain
symmetries that are recovered in a statistical sense. Following
Kolmogorov's theory, the small scales can be uniquely described by
simple parameters such as the kinematic viscosity $\nu$ of the fluid
and the mean dissipation rate.  However, numerous experimental and
numerical studies have reported a substantial deviation from
Kolmogorov's prediction \cite{frisch1995,sreenivasan1996}. The
turbulent fluid motion is extremely fluctuating and the intensity of
these fluctuations increases strikingly with increasing Reynolds
number or decreasing scale. This phenomenon is referred to as internal
intermittency. 
\citet{nelkin1994universality} claimed that the origin of
intermittency lies in the non-linear and non-local vortex stretching
mechanism. The consequence of internal intermittency is the break-down
of small-scale universality, which dramatically complicates
theoretical approaches from first principles. Even for the most
canonical flows, such as homogeneous isotropic turbulence, no closed
theory for intermittency exists \cite{peters2016higher}.

It is customary to investigate the statistical structure of turbulence
by means of velocity or scalar increments over a given range of
scales. The scalar can be either concentration or temperature,
provided that buoyancy effects are negligible. Kolmogorov's scaling
theory has been generalized to scalar fields by
\citet{obukhov1949temperature} and \citet{corrsin1951spectrum}
(referred to as KOC-theory in the following). The moments of the
increments are known as structure functions. Transport equations for
the structure functions have been first derived by
\citet{kolmogorov1941} for the velocity field and by
\citet{yaglom1949} for a scalar field. Physically, structure functions
provide an information about the energy at a given length-scale $r$
and all smaller scales \cite{davidson2004}. In other words, structure
functions at scale $r$ are contaminated with information coming from
all smaller length-scales. Spectral representations have the same
deficiency, as structure functions and energy spectra are related by
an integral transformation. Moreover, by defining structure functions
over a given scale $r$, the information of the length-scale
distribution of the turbulent field is lost after applying an
ensemble-average operation.

The concept of self-similarity is less strict than universality and
has been one of the key elements in shaping our understanding of
turbulent flows \cite{speziale1992energy,george1992decay}. The
self-similarity hypotheses was first put forward by
\citet{de1938statistical} for the correlation functions of the
velocity and afterwards, several other problems in turbulence research
have been approached in this framework
\cite{antonia2003similarity,tang2016complete,danaila2017self,djenidi2017self}. The
assumption of self-similarity imposes certain constrains on the
dynamics of the flow. In particular, complete self-similarity requires
that all statistics, such as structure functions or correlation
functions, can be expressed by functional forms. For instance, the
second-order scalar structure function $\avg{(\delta\phi)^2}$ can be
written as
\begin{equation}
  \avg{(\delta \phi)^2} = A(t) f(\tilde r) \,,
\end{equation}
where the pre-factor $A(t)$ depends solely on time, while the
normalized shape function $f(\tilde r)$ depends solely on the
normalized separation distance $\tilde r$, with $\tilde r = r/L(t)$,
and $L(t)$ being a single characteristic length-scale.  However, many
flows do not satisfy complete self-similarity
\cite{meldi2013further}. The term partial self-similarity refers to
flows in which self-similarity is valid for a restricted range of
scales only, requiring at least two different length-scales to
describe statistical quantities. Determining these scales and its
scaling is of high relevance for understanding and modeling
turbulence.

A novel statistical approach to investigate the local structure of
turbulence was proposed by \citet{wang2006length,wang2008length} by
exploiting the topological features of local extremal points in
turbulent fields. The motivation behind this approach is the fact that
the extremal points inherit physics from the dynamics of turbulence:
fluctuations disturb the turbulent field and create new extremal
points, while diffusion will smooth the field and annihilate the
extremal points. Strain acting on the turbulent field does not change
the number of extremal points but can move the position of the
extremal points relatively to each other.  The extremal points can be
considered within different frameworks, such as along straight lines
\cite{gauding2015line}, gradient trajectories \cite{wang2006length},
stream lines \cite{wang2010properties}, vortex lines
\cite{wang2013new}, and Lagrangian trajectory paths
\cite{wang2014analysis}. Recently, statistics of extremal points have
been studied in the context of turbulent combustion
\cite{gauding2017dissipation,chakraborty2014streamline} 
  and within a framework to predict local detonation events in
  super-charged spark-ignition engines \cite{peters2013super}.

\begin{figure}
  \centering
  \includegraphics[width=0.66\linewidth]{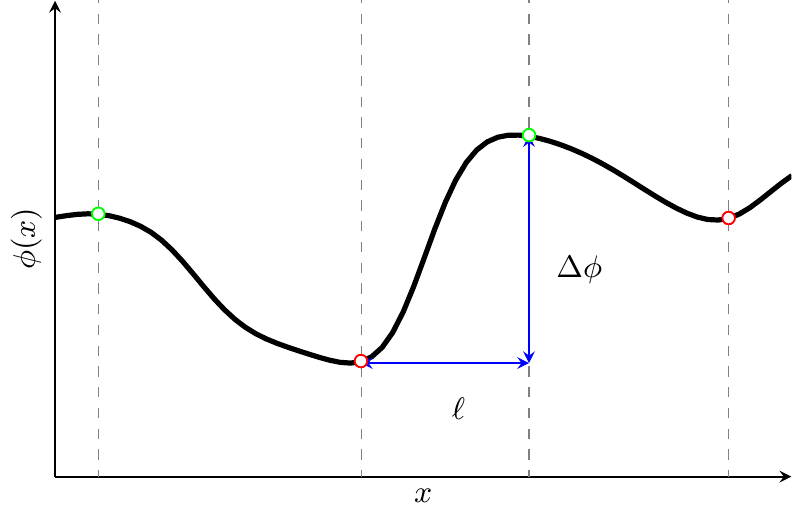}
  \caption{Illustration of the parameterization of line segments by
    $\Delta\phi$ and $\ell$ for a signal $\phi(x)$. Line segments are
    bounded by adjacent local minimum and maximum points, which are
    indicated by red and green circles, respectively.}
  \label{fig:line}
\end{figure}

In this work, we adopt the approach of \citet{gauding2015line} to
analyze turbulent scalar mixing in decaying homogeneous isotropic
turbulence.  The turbulent signal of the passive scalar
$\phi(\vec x,t)$ is decomposed along a straight line into piece-wise
monotonously increasing or decreasing segments. These so-called line
segments start at a local minimum point of the scalar field and end at
a local maximum point of the scalar field or vice versa.  By this
definition, line segments can be parameterized by the distance
$\ell=x_{\rm end} - x_{\rm start}$ and the scalar difference
$\Delta \phi=\phi(x_{\rm end})-\phi(x_{\rm start})$ between the end
and start point.  This concept is demonstrated in
fig.~\ref{fig:line}. Depending on the sign of $\Delta\phi$, there are
positive (increasing) or negative (decreasing) segments.  The
decomposition by line segments is self-contained and has the
properties of completeness and uniqueness meaning that each material
point is included once and only once in the decomposed object.

 Such decomposition is physically meaningful in the
  following sense. By decomposing the turbulent field into
  space-filling sub-units, the properties of the entire field can be
  reproduced from the statistics of the relatively simple
  sub-units. In this sense, the complexity of the problem will be
  reduced. Specifically, suppose the sub-units can be characterized by
  a set of representative parameters $(p_1, \cdots, p_n)$. In
  principle it will be much more challenging to parameterize a
  turbulent quantity $X$ for the entire field than to parameterize $X$
  with $(p_1, \cdots, p_n)$ for each individual sub-unit because of
  the relatively simple sub-unit structure. When the joint probability
  density function $P(p_1,\cdots, p_n)$ of the parameter set $p_i$ is
  known, the ensemble average of $X$ is determined as
  \begin{equation}
    \label{eq:X}
    \avg{X} = \int \cdots \int X(p_1, \cdots, p_n) P(p_1, \cdots, p_n)
    {\rm d} p_1 \cdots {\rm d} p_n \,.
  \end{equation}
  It is worthy noting that eq.~\eqref{eq:X} for reconstructing $X$ is
  valid if and only if the sub-units are space-filling.

Finally, we want to emphasize the difference between the scalar
difference $\Delta \phi$ of line segments and the increment
$\delta \phi = \phi(\vec x + \vec r) - \phi(\vec x)$ used in classical
theories. The increment $\delta \phi$ is computed continuously along a
straight line and the separation distance $r$ between the two points
is imposed externally.  For line segments, the length-scale $\ell$
results from the turbulent field itself, rather than being prescribed
externally, and the scalar difference $\Delta \phi$ is conditioned on
local extremal points.  This approach is therefore capable to
characterize the local structure of turbulence.

The objective of this study is to use the concept of line segments to
analyze scalar mixing in decaying homogeneous isotropic
turbulence. The remainder of the paper is as follows. In
sec.~\ref{sec:dns} we present the direct numerical simulations (DNS)
on which the analysis is based. In sec.~\ref{sec:length} we discuss
the length distribution of line segments and motivate the scaling of
the mean length by using a statistical approach. Joint statistics of
$\Delta\phi$ and $\ell$ and their self-similarity are presented in
sec.~\ref{sec:joint}. Conditional averages of these quantities are
presented in sec.~\ref{sec:cond}. The statistics are discussed from a
self-similarity perspective. We identify quantities that satisfy
complete self-similarity, such as the distribution of the length
$\ell$, and quantities that reveal only partial self-similarity, such
as the joint distribution of $\Delta\phi$ and $\ell$.  We conclude the
paper in sec.~\ref{sec:conclusion}.

\section{Direct numerical simulations}
\label{sec:dns}
Highly resolved direct numerical simulation (DNS) of decaying
homogeneous isotropic turbulence has been carried out.  The DNS solves
the incompressible Navier-Stokes equations by a pseudo-spectral method
in a triply periodic cubic box with size $2 \pi$. Similar to the
approach by \citet{mansour1994decay}, the Navier-Stokes equations are
formulated in spectral space as
\begin{equation}
  \label{eq:ns}
  \pp{}{t} \left( \hat u_i \exp(\nu \kappa^2 t) \right) = \exp(\nu
  \kappa^2 t) P_{ij} \hat H_j \,,
\end{equation}
where
\begin{equation}
\hat H_j   = - i \kappa_i \mathcal F \left(  u_i u_j \right) 
\end{equation} 
is the Fourier transform of the non-linear term and
$P_{ij} = \delta_{ij} - \kappa_i \kappa_j/\kappa^2$ is the projection
operator that imposes incompressibility. Einstein's summation
convention is used, which implies summation over indices appearing
twice. In eq.~\eqref{eq:ns}, the wave-number vector is denoted by
$\vec \kappa$, the Fourier transform of the velocity field is denoted
by $\hat {\vec u}$, and $\nu$ is the kinematic viscosity.  An
integrating factor technique is used for an exact integration of the
viscous term. Temporal integration is carried out by a low-storage,
stability preserving, third-order Runge-Kutta scheme. The non-linear
term is computed in physical space and a truncation technique with a
smooth spectral filter is applied to reduce aliasing errors, see
\citet{hou2007computing}.  The library P3DFFT has been used for
spatial decomposition and to perform the fast Fourier transform
\cite{pekurovsky2012p3dfft}. The code employs a hybrid MPI/OpenMP
parallelization. The simulations used more than 1.8 Million concurrent
threads and have been carried out on the super-computer JUQUEEN at
research center Juelich \cite{stephan2015juqueen}.  More details about
the numerical procedure and the parallelization strategy are given by
\citet{gauding2015line,gauding2017high} and
\citet{goebbert2016direct,goebbert2016overlapping}.

A necessary constraint that has to be satisfied by the DNS is an
adequate resolution of all relevant scales. For the specific case of
decaying turbulence this requires resolving the smallest scales down
to the viscous cut-off scale, while keeping the integral length-scale
$l_t$, defined as
\begin{equation}
  l_t = \frac{3 \pi}{4} \frac{\int \kappa^{-1} E(\kappa) {\rm d}
    \kappa}
  {\int E(\kappa) {\rm d}\kappa} \,,
\end{equation}
small compared to the size of the computational domain to reduce
confinement effects.  Following \citet{mansour1994decay}, we require
that the resolution condition $\kappa_{\rm max} \eta \ge 1$ is
satisfied for all times, where $\eta$ is the Kolmogorov length-scale
and $\kappa_{\rm max}$ is the largest resolved wave-number. A grid
resolution of $4096^3$ points is used to appropriately account for all
relevant length-scales.

The initial velocity field is generated in spectral space to be random
and statistically isotropic.  It satisfies incompressibility and obeys
a prescribed energy spectrum of the Batchelor-Proudman type
\cite{batchelor1956large}, i.e.
\begin{equation}
  \label{eq:spec}
  E(\kappa,0) \propto \kappa^4 \exp \bigg(-2  \Big(
  \frac{\kappa}{\kappa_p} \Big)^2 \bigg) \,.
\end{equation}
In eq.~\eqref{eq:spec}, $\kappa_p=15$ is the wave-number at which the
maximum of the initial spectrum $E(\kappa,0)$ occurs. The chosen value
of $\kappa_p$ is a compromise between limiting the confinement effect
imposed through the finite size of the computational domain and the
goal of reaching a high Reynolds number. Following
\citet{ishida2006decay}, the initial state of freely decaying
turbulence may be described by a Reynolds number defined as
\begin{equation}
  \mathit{Re}_0 = \frac{u'_0}{\kappa_p \nu} \,.
\end{equation}
With a kinematic viscosity of $\nu=1.82 \cdot 10^{-4}$, and an initial
turbulence intensity $u'_0=2.58$, where
$u'^2 = \frac{2}{3} \avg{k}=\frac{1}{3} \avg{u_i^2}$, we obtain an
initial Reynolds number $\mathit{Re}_0$ of 945.  Further details of
the initialization of the simulation are presented in
table~\ref{tab:dns1}.

\begin{figure}
  \centering
  \includegraphics[width=0.66\linewidth]{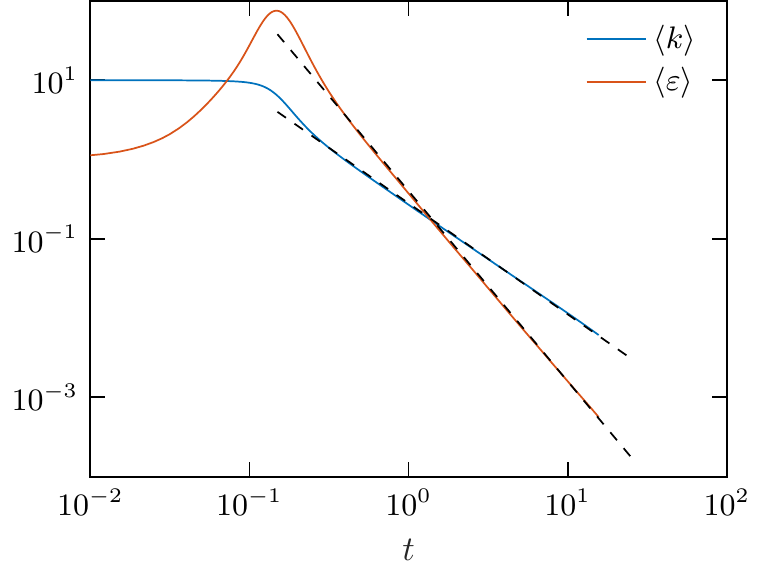}
  \caption{Temporal evolution of $\avg{k}$ and $\avg{\varepsilon}$
    obtained from DNS. The black dashed lines refer to the analytical
    scaling exponents of $-10/7$ and $-17/7$, respectively.}
  \label{fig:dns1}
\end{figure}

Prescribing a $\kappa^4$ energy spectrum at the small wave-numbers is
equivalent to a conservation of Loitsyansky's integral
\cite{davidson2004,rotta2010turbulente} and requires
\begin{equation}
  \label{eq:loy}
  \avg{k} l_t^5 = \mathrm{constant} \,.
\end{equation}
Equation \eqref{eq:loy} implies a temporal decay of the mean turbulent
energy and the mean dissipation as $\avg{k} \propto t^{-10/7}$ and
$\avg{\varepsilon}\propto t^{-17/7}$, respectively. The DNS recovers
these scaling laws after an initial transient for $t>0.3$ over nearly
two decades, cf.~fig.~\ref{fig:dns1}. In the following, we refer to
this range as the self-similar decay. However, it is important to note
that Batchelor-Proudman turbulence with $E(\kappa) \propto \kappa^4$
is not completely self-similar. Complete self-similarity requires a
constant Reynolds number, which is possible in decaying homogeneous
isotropic turbulence only under the condition that the turbulent
kinetic energy and the energy dissipation decay as
$\avg{k} \propto t^{-1}$, and $\avg{\varepsilon} \propto t^{-2}$,
respectively, see \citet{ristorcelli2006passive,ristorcelli2003self}
and references therein.

\begin{table}
  \centering
  \caption{Initial properties of the DNS. 
  }
  \begin{tabular}{ll}
    \toprule
    Grid size $N^3$ & $4096^3$ \\
    Peak wave-number $\kappa_p$ & 15 \\
    Viscosity $\nu$ & $1.82 \cdot 10^{-4}$ \\
    Reynolds number $u_0'/(\nu \kappa_p) $ & 945 \\
    Turbulence intensity $u_0' $ & 2.58 \\
    Schmidt number $\mathit{Sc}$ & 1 \\
    Integral length-scale $l_t$ & 0.2 \\
    \bottomrule
  \end{tabular}
  \label{tab:dns1}
\end{table}

For the analysis of turbulent mixing, an additional
advection-diffusion equation for the scalar fluctuations
$\phi(\vec x,t)$ is solved, i.e.
\begin{equation}
  \label{eq:dns2}
  \pp{\phi}{t} + u_i \pp{\phi}{x_i} = D \pp{^2 \phi}{x_i^2} -  \Gamma
  u_2  \,,
\end{equation}
where $D$ is the molecular diffusivity. In the following, we consider
a unity Schmidt number meaning that the kinematic viscosity $\nu$
equals the molecular diffusivity $D$.  A uniform mean scalar gradient
$\Gamma$ is imposed on the scalar field. In this configuration, the
velocity field is decaying while the scalar field is subject to a
continuous injection of energy at the large scales. A similar flow
configuration was studied experimentally by \citet{bahri2015self} and
\citet{warhaft2000passive}. The scalar field is initialized with
delta-correlated fluctuations at low intensity that are initially
uncorrelated with the velocity field, i.e. $\avg{u_i\phi}=0$. By this
approach, scalar structures develop naturally from the non-linear
coupling to the velocity field and the injection of energy by the mean
scalar gradient $\Gamma$.

\begin{figure}
  \centering
  \includegraphics[width=0.66\linewidth]{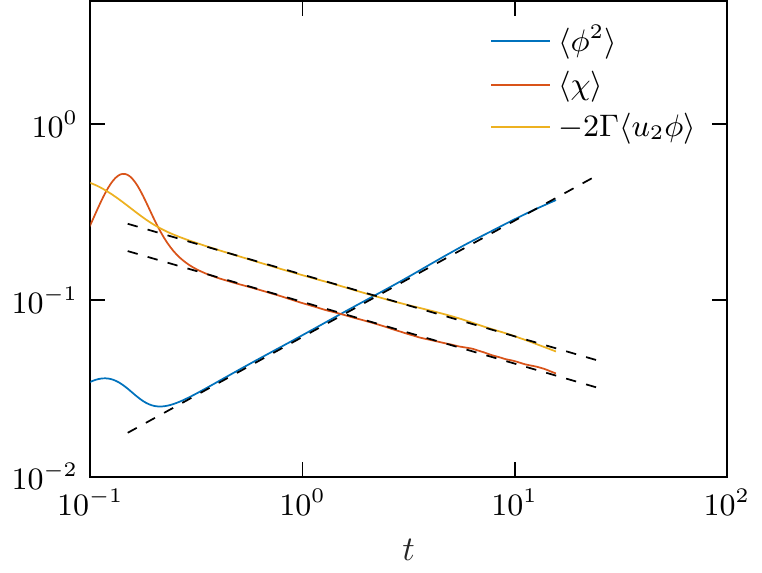}
  \caption{Temporal evolution of the scalar variance $\avg{\phi^2}$
    and its production $- 2 \Gamma \avg{u_2 \phi}$ and dissipation
    $\avg{\chi}$. The black dashed lines are displayed as a visual
    guide to indicate power-law scaling.}
  \label{fig:dns2}
\end{figure}

The scalar variance $\avg{\phi^2}$ is governed by the evolution
equation
\begin{equation}
  \label{eq:phi2}
  \pp{\avg{\phi^2}}{t} = - 2 \Gamma \avg{u_2 \phi} - \avg{\chi} \,,
\end{equation}
where $\avg{\chi}$ is the mean scalar dissipation, defined as
\begin{equation}
  \avg{\chi} = 2D \avg{\left( \pp{\phi}{x_i} \right)^2} \,.
\end{equation}
Equation \eqref{eq:phi2} shows that the change of the scalar variance
is determined by a balance between production
$- 2 \Gamma \avg{u_2 \phi}$ and dissipation $\avg{\chi}$ of scalar
energy, see fig.~\ref{fig:dns2} for a temporal evolution of these
terms. After the initial transient, production of scalar energy
exceeds scalar dissipation resulting in a temporal increases of
$\avg{\phi^2}$. We note in passing that all terms in
eq.~\eqref{eq:phi2} reveal after the initial transient a power-law
scaling.

\begin{figure}
  \centering
  
  \includegraphics[width=0.66\linewidth]{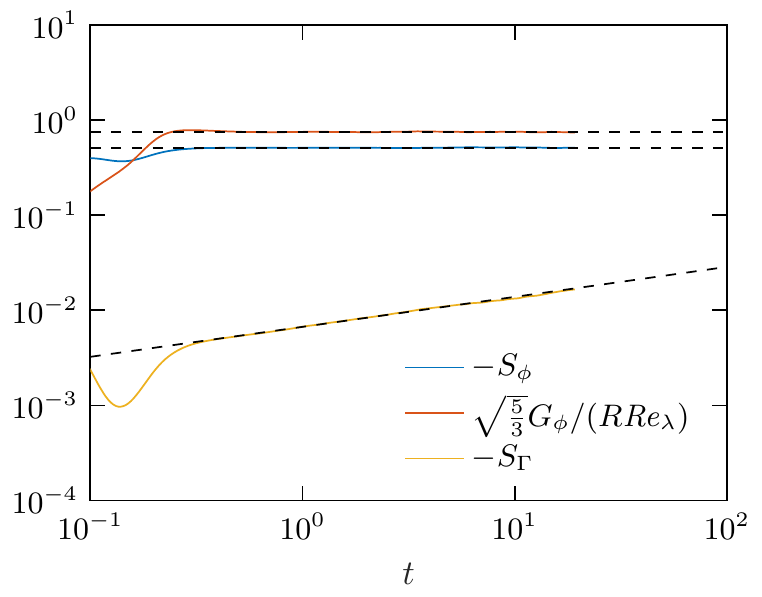} 
  \caption{Temporal evolution of the normalized production of scalar
    dissipation by vortex stretching $-S_\phi$ and mean scalar
    gradient $-S_\Gamma$, and normalized destruction of scalar
    dissipation by diffusion
    $\sqrt\frac{5}{3} G_\phi/(R\mathit{Re}_\lambda)$.  The black
    dashed lines are displayed as a visual guide. \label{fig:sdt}}
\end{figure}

To further characterize the DNS, it is meaningful to analyze the
transport equation for the mean scalar dissipation, which can be
written as
\begin{equation}
  \label{eq:gs}
  \frac{1}{\avg{\chi}} \dd{\avg{\chi}}{t} = -4
  \sqrt{\frac{\avg{\varepsilon}}{\nu}}
  \left(
    (S_\phi + S_\Gamma) 
    + \sqrt{\frac{5}{3}}\frac{G_\phi}{\mathit{Re}_\lambda R}
  \right) \,.
\end{equation}
Equation \eqref{eq:gs} generalizes the derivation by
\citet{gonzalez1998approach}, \citet{Zhou2000}, and
\citet{danaila2001effect} to statistically homogeneous but anisotropic
turbulence with large-scale production. In eq.~\eqref{eq:gs}, we
introduced the mixed velocity-scalar gradient skewness
\begin{equation}
  \label{eq:sphi}
  S_\phi = \frac{\avg{g_i g_j A_{ij}}}{\avg{g^2} \avg{A_{kl}^2}^{1/2}} \,,
\end{equation}
the scalar gradient destruction coefficient
\begin{equation}
  \label{eq:gphi}
  G_\phi= \frac
  {\avg{\phi^2} }
  {\avg{g^2}^2} \avg{\left( \pp{^2 \phi}{x_i \partial x_k}  \right)^2}
  \,,
\end{equation}
and the normalized scalar gradient production due to the mean gradient
$\Gamma$
\begin{equation}
  \label{eq:sgamma}
  S_\Gamma =\frac
  {\Gamma \avg{ g_k A_{2k} }}
  {\avg{g^2} \avg{A_{ij}^2}^{1/2}} \,.
\end{equation}
The velocity gradient tensor is given by
$A_{ij}=\partial u_i /\partial x_j$, the scalar gradient is denoted by
$g_i=\partial \phi/\partial x_i$, and the velocity-scalar time-scale
ratio is defined as
\begin{equation}
 R = \frac{\avg{\phi^2}}{\avg{\chi}}
 \frac{\avg{\varepsilon}}{\avg{u_i^2}} \,.
\end{equation}
During the self-similar decay, the velocity-scalar time-scale ratio
$R$ is virtually constant and equals approximately 1.1 for the present
DNS.  Equation \eqref{eq:gs} contains two different production
mechanisms for the mean scalar dissipation: small-scale production due
to the vortex-stretching mechanism of turbulence
\cite{brethouwer2003micro} described by $-S_\phi$, and large-scale
production due to the mean scalar gradient $\Gamma$ covered by the
term $-S_\Gamma.$ Destruction of scalar dissipation results from
molecular diffusivity and is described by $G_\phi$.
Figure~\ref{fig:sdt} displays the temporal evolution of the different
terms. Both terms, $- S_\phi$ and
$\sqrt{5/3}G_\phi/(R \mathit{Re}_\lambda)$, tend to a constant, where
the destruction $\sqrt{5/3}G_\phi/(R \mathit{Re}_\lambda)$ prevails
over the production $-S_\phi$.  The production of scalar dissipation
by the mean scalar gradient $-S_{\Gamma}$ increases by following a
power-law with time, but remains negligible compared to production by
vortex stretching $-S_\phi$.  The contribution of the large-scale
production $-S_\Gamma$ to the budget in eq.~\eqref{eq:gs} gains
importance during the self-similar decay. This results from a decrease
of the Reynolds number, which leads to a reduced scale separation
between small and large scales.

\begin{figure}
   \centering
   \includegraphics[width=0.66\linewidth]{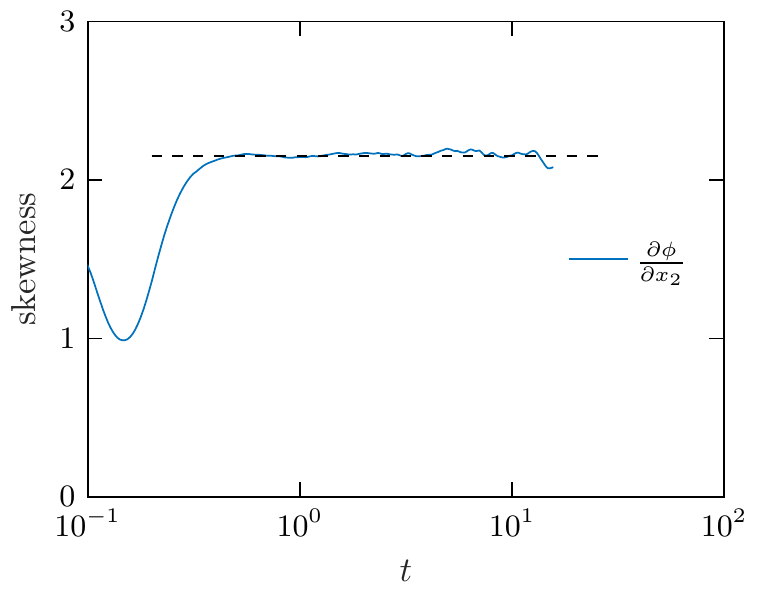}
   \caption{ Temporal evolution of the scalar gradient skewness
     $S$ in $x_2$-direction (parallel to the direction of the scalar
     mean gradient). The black dashed line refers to a constant value
     of 2.16.  The scalar gradient skewness perpendicular to the
     scalar mean gradient (not shown) is zero.}
   \label{fig:skew}
 \end{figure}

 Under the influence of a mean scalar gradient, statistics of
  the scalar field are characterized by an anisotropy, such as a
  non-zero scalar gradient skewness, i.e.
  \begin{equation}
    S = \frac{\avg{\left( \pp{\phi}{x_2} \right)^3}}
    {\avg{\left( \pp{\phi}{x_2} \right)^2}^{3/2}} \,,
  \end{equation}
  in the direction of the mean scalar gradient
  \cite{sreenivasan1980skewness,tong1994passive,bos2014anisotropy}. The
  non-zero skewness originates from a preferential alignment of
  coherent scalar structures with the direction of the mean scalar
  gradient, which manifest themselves as cliffs and ramps, i.e.\ steep
  gradients followed by relatively well mixed regions
  \cite{brethouwer2003micro}. Figure~\ref{fig:skew} shows the temporal
  evolution of the scalar gradient skewness~$S$. During the
  self-similar decay,~$S$ approaches a nearly constant value close
  to~2.16. The scalar signal in planes perpendicular to the scalar
  mean gradient does not exhibit a statistical asymmetry, and the
  scalar gradient skewness in these directions is zero.


\begin{figure}
  \centering
  \includegraphics[width=0.62\linewidth]{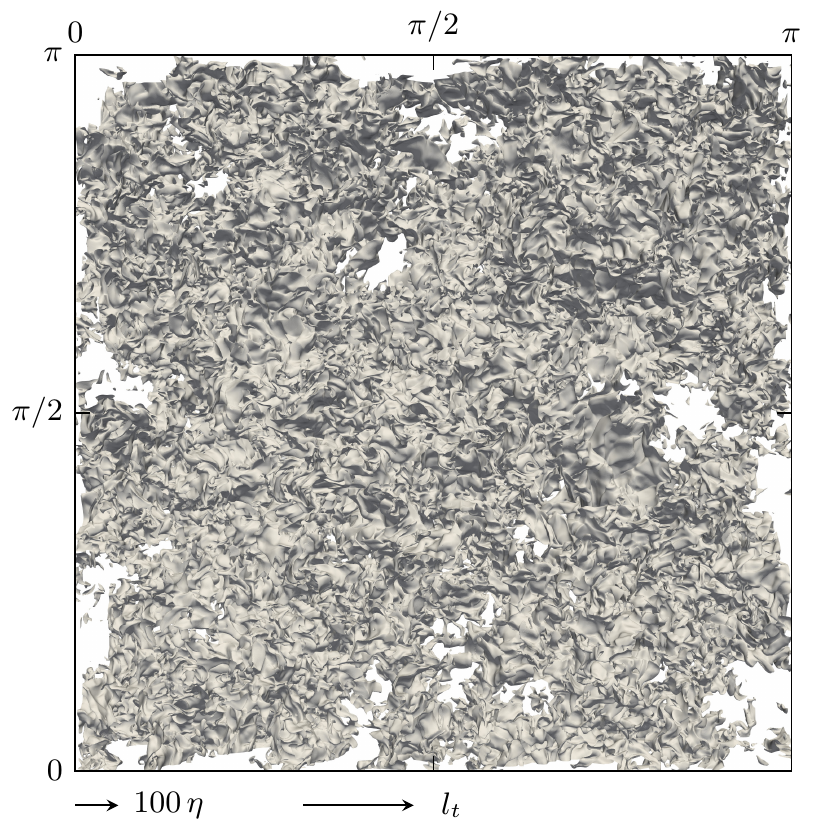} \\[1.5ex] 
  \includegraphics[width=0.62\linewidth]{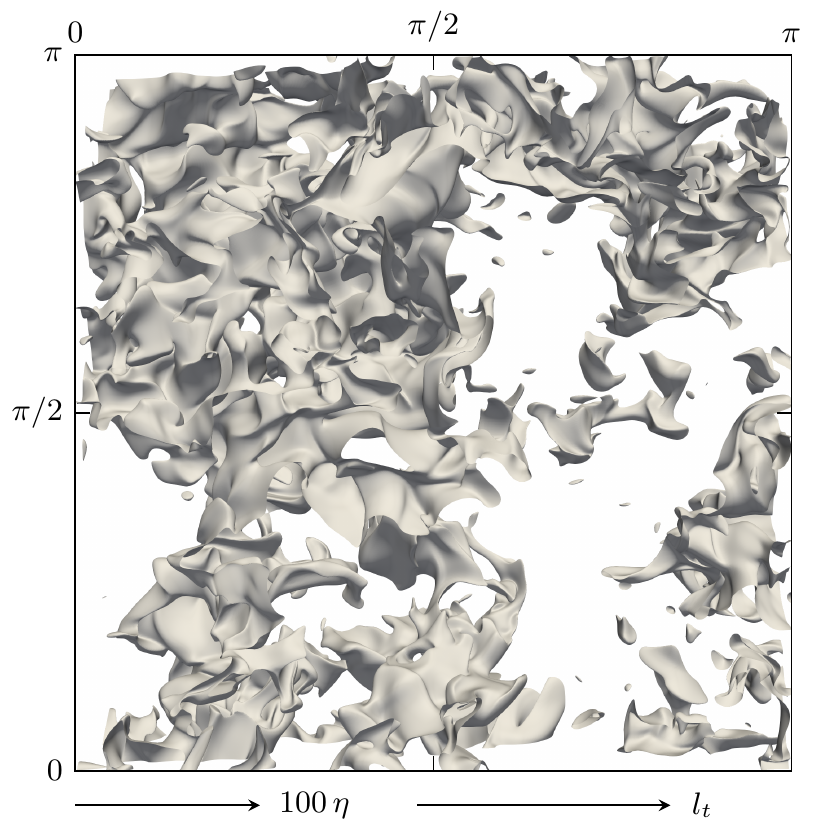}
  \caption{ Visualization of the iso-scalar surface
    obtained from case D2 (top) and case D6. Only a subset of the
    computational domain of size $\pi \times \pi \times \pi/4$ is
    shown for clarity. The length of the arrows represents 100 times
    the Kolmogorv length scale $\eta$ and the integral length scale
    $l_t$, respectively. }
  \label{fig:vis}
\end{figure}

 Characteristic properties of the DNS are summarized in
 table~\ref{tab:dns2} for six different times steps (denoted by D1-D6)
 that are used for further analysis.  During that time, the
 Taylor-based Reynolds number decreases by nearly a factor of 2 from
 95.8 to 54.6. Ensemble-averages (denoted by angular brackets) are
 computed by virtue of homogeneity over all grid points in the
 computational domain, and additionally, to improve accuracy of
 statistical quantities, over three statistically independent
 realizations of the DNS. With this procedure, at each time step,
 statistics are computed over more than 200 Billion grid
 points.  Figure \ref{fig:vis} displays a visualization of
   the scalar iso-surface for two different time steps. At the early
   time step (D2), the scalar iso-surface is strongly twisted and
   folded and reveals regions of high curvature. During the
   self-similar decay, the iso-surfaced is smoothed and larger
   coherent structures are visible.

\begin{table}
  \centering
  \caption{Characteristic properties of the DNS at different $t$}
  \begin{tabular}{lcccccc}
    \toprule  & D1 & D2 & D3 & D4 & D5 &D6 \\ \midrule
    $t$ & 0.5 & 0.9 & 2 & 4 & 7 & 10.3 \\
    $\avg{k}$  & 0.7468 & 0.3097 & 0.1034 & 0.0402 & 0.0187 & 0.0109 \\
    $\avg{\varepsilon}$  & 2.2243& 0.4768& 0.0705& 0.0137& 0.0037&
                                                                   0.0015
    \\
    $\mathit{Re}_\lambda$  & 95.8 & 85.8 & 74.5 & 65.7 & 59.1 & 54.6 \\
    $l_t$  & 0.38 & 0.47 & 0.63 & 0.79 & 0.95 & 1.05 \\
    $\avg{\phi^2}$ & 0.0388 & 0.0587 & 0.0993 & 0.1590 & 0.2330 &
                                                                  0.3036
    \\
    $\avg{\chi}$  & 0.1246 & 0.0998 & 0.0753 & 0.0593 & 0.0500 & 0.0443
    \\
    $-2 \Gamma \avg{u_2 \phi}$    & 0.0897& 0.0714 &  0.0541& 0.0435&
                                                                      0.0363
                                       & 0.0317 \\
    $R $    & 1.07& 1.10 & 1.11 & 1.09 & 1.09 & 1.09 \\ 
    $S$    &2.15  &   2.15 &    2.17&    2.17&    2.16&    2.15 \\
    $\kappa_{\rm max} \eta$   & 2.15& 3.16& 5.10 & 7.69 & 10.69 & 13.45
    \\ \bottomrule
  \end{tabular}
  \label{tab:dns2}
\end{table}

Finding the local extremal points of the signal $\phi$ along a
straight line in direction $x_1$ turns into the problem of finding the
roots of its first derivative $\phi_x=\partial \phi /\partial
x_1$. The derivatives of $\phi$ are calculated by a spectral method
and exactly interpolated to a finer mesh. On this new mesh, the
zero-crossings of the first derivative act as the starting points for
a Newton iteration that yields the exact position of the extremal
points. Detecting local extremal points to define line segments
requires a sufficiently smooth and well resolved turbulent field. In
order to appropriately identify line segments, we require that the
turbulent scalar field can be locally expanded as a Taylor series up
to order two. This is equivalent to the condition that the turbulent
scalar field is locally two times continuously differentiable. If this
condition is violated, statistics of line segments are
under-resolved. From parameter studies, we found
$\kappa_{\rm max} \eta \ge 2.5$ to be a necessary
condition. Therefore, we report results for line segments for
$t\ge 0.9$.

 In the following, we present the analysis of the DNS by line
  segments with a special emphasizes on the question whether
  statistics of line segments are self-similar.  Line segments are
  computed in $x_1$-direction, which is perpendicular to the mean
  scalar gradient. The effect of anisotropy is discussed in
  \ref{sec:aniso}.

\section{The length distribution of line segments}

 \label{sec:length}

\begin{figure}
   \centering
   \includegraphics[width=0.66\linewidth]{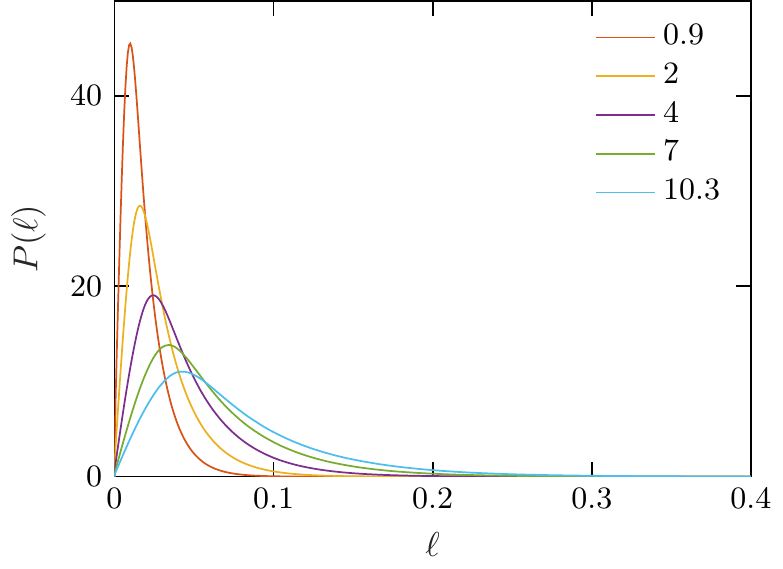}
   \includegraphics[width=0.66\linewidth]{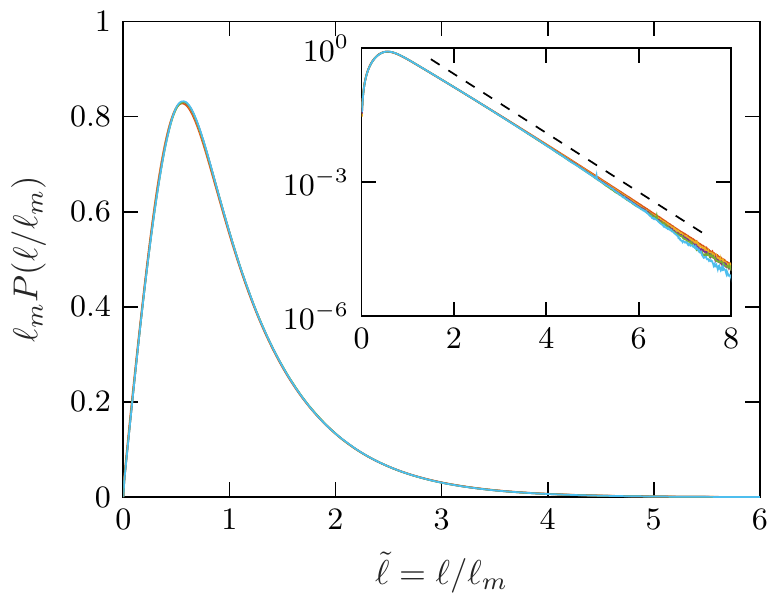}
   \caption{Length distribution $P(\ell)$ of line segments (top), and
     normalized length distribution
     $\tilde P(\tilde \ell) = \ell_m P(\ell/\ell_m)$ (bottom) for
     different time steps. The black dashed line in the inset
     indicates an exponential tail, i.e.\ $\tilde P(\tilde \ell) \propto
     \exp(-1.5 \, \tilde \ell)$.}
   \label{fig:pdf_ell}
 \end{figure}

 Two characteristic parameters, the length $\ell$ and the scalar
 difference $\Delta\phi$ have been defined to characterize line
 segments. The linear length~$\ell$ between adjacent extremal points
 provides a measure for the length-scales present in turbulent
 fields. 
 Figure~\ref{fig:pdf_ell} (top) shows the length-distribution
 $P(\ell)$ for different time steps during the decay. It is observed
 that the curves do not collapse and that for later times, the maximum
 of $P(\ell)$ is shifted towards larger scales. In other words: the
 mean length $\ell_m$ of dissipation elements, which is related to
 $P(\ell)$ by
 \begin{equation}
   \ell_m= \int_0^\infty \ell P(\ell) {\rm d} \ell \,,
 \end{equation}
 increases during the decay. This finding is in agreement with the
 increase of all other turbulent length-scales in decaying turbulence.
 
 Figure~\ref{fig:pdf_ell} (bottom) shows the probability density
 function (pdf) of $\ell$, rescaled with the respective mean length
 $\ell_m$ for each time step.  By normalization with a single
 parameter, namely, the mean length $\ell_m$, the pdfs of the length
 collapse to a single curve.  This finding indicates that the pdf of
 the non-dimensional length $\tilde{\ell}=\ell/\ell_m$, which can be
 calculated as
 \begin{equation}
   \label{eq:pell}
   \tilde P(\tilde \ell) = \ell_m P(\ell/\ell_m)
 \end{equation}
 is completely self-similar in decaying turbulence.  It is worth
 mentioning that self-similarity of $\tilde P(\tilde \ell)$ was
 observed before in homogeneous isotropic forced turbulence for a
 broad range of the Reynolds numbers between $\mathit{Re}_\lambda=88$
 and $\mathit{Re}_\lambda=529$, cf.~\citet{gauding2015line}.  By
 definition, $\tilde P(\tilde \ell)$ satisfies two constrains, i.e.
 \begin{equation}
   \int_0^\infty \tilde P(\tilde \ell) {\rm d} \tilde \ell = 1 \,,
 \end{equation}
 and
 \begin{equation}
   \int_0^\infty \tilde \ell \tilde P(\tilde \ell) {\rm d} \tilde \ell = 1 \,.
 \end{equation}

 As displayed in fig.~\ref{fig:pdf_ell}, $\tilde P(\tilde \ell)$
 declines steeply towards the origin as small segments are annihilated
 by molecular diffusion. For larger length-scales the pdf exhibits a
 clear exponential decay indicating that these segments are governed
 by a Poisson process (see inset of fig.~\ref{fig:pdf_ell}). This is
 plausible as for larger length-scales the probability of the
 occurrence of an extremal point is independent of the adjacent
 ones. 
 The non-gaussianity of $\tilde P(\tilde \ell)$ is a feature that is
 shared with the pdf of other turbulent quantities, such as the pdf of
 the scalar gradient or scalar increment
 \cite{warhaft2000passive,pumir1994numerical,holzer1994turbulent}. However,
 different from these quantities, $\tilde P(\tilde \ell)$ reveals
 complete self-similarity under rescaling with a single length-scale.
 The pdfs of scalar gradients or increments cannot be superposed by a
 non-trivial rescaling procedure, because of the existence of
 stretched exponential tails, which depend on either Reynolds number,
 length-scale or other non-universal features. For comparison, further
 statistics of the scalar gradient will be presented in
 sec.~\ref{sec:joint}.

\begin{figure}
   \centering
   \includegraphics[width=0.66\linewidth]{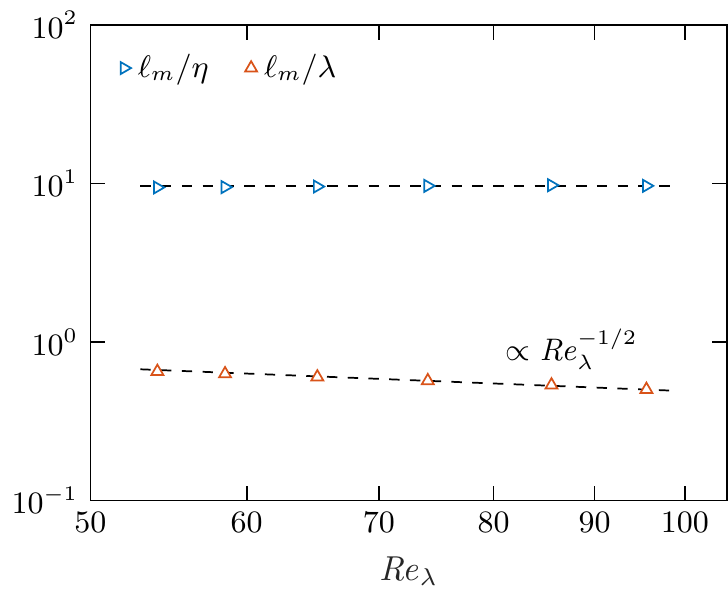}
   \caption{Scaling of the mean length $\ell_m$ of line segments with
     the Kolmogorov length-scale $\eta$ and the Taylor micro-scale
     $\lambda$ as a function of $\mathit{Re}_\lambda$. }
   \label{fig:ell_time}
 \end{figure}

 DNS suggests that the mean length $\ell_m$ is the only characteristic
 length-scale for the normalized distribution function
 $\tilde P( \tilde \ell)$. To obtain the dimensional pdf $P(\ell)$
 from the normalized pdf $\tilde P(\tilde \ell)$, the scaling of the
 normalization quantity, i.e.\ the mean length $\ell_m$, is
 required. This information is provided by fig.~\ref{fig:ell_time},
 where the normalized mean length $\ell_m/\eta$ is shown as the
 function of the Taylor-based Reynolds number. The DNS results
 indicate that the mean length $\ell_m$ scales with the Kolmogorov
 length-scale, i.e.\
 \begin{equation}
   \label{eq:lm_eta}
   \ell_m \approx 10 \eta \,.
 \end{equation}
  Equation~\eqref{eq:lm_eta} is well satisfied for scalar fields
   in decaying turbulence and forced turbulence \cite{gauding2015line}
   for a wide range of different Reynolds numbers. However, the
   proportionality constant given in eq.~\eqref{eq:lm_eta} may not be
   universal and may depend on the large-scales of the flow. Moreover,
   the proportionality constant is flow-field dependent and is, for
   example, different for the signal of the longitudinal or
   transversal velocity components. In the next paragraph, we provide
   a justification for the scaling relation given by
   eq.~\eqref{eq:lm_eta} based on a statistical theory developed by
   \citet{rice1944mathematical}.


\citet{rice1944mathematical} proved that for any homogeneous Gaussian
stochastic process $\phi(x)$, the number of zero-crossings is given by
 \begin{equation}
   \label{eq:rice}
   N_0 = \frac{1}{\pi} \left[ - \frac{f''_\phi(0)}{f_\phi(0)}
   \right]^{1/2} \,.
 \end{equation}
 In eq.~\eqref{eq:rice}, $f_\phi$ is the normalized correlation function
 of the signal $\phi(x)$, defined as
 \begin{equation}
   f_\phi(r) = \frac{\avg{\phi(x)\phi(x+r)}}{\avg{\phi^2}} \,,
 \end{equation}
 and $f''_\phi(r)$ is the second derivative of $f_\phi(r)$. By virtue
 of homogeneity, the correlation functions depend only on the
 separation distance $r$. \citet{liepmann1949anwendung} and
 \citet{sreenivasan1983zero} found from experiments that
 eq.~\eqref{eq:rice} is valid for the turbulent velocity field,
 despite the fact that turbulence is clearly not
 Gaussian. \citet{liepmann1949anwendung} noticed that for the validity
 of eq.~\eqref{eq:rice}, essential statistical independence between
 $\phi(x)$ and its first derivative $\phi_x(x)$ is required.

 This observation encouraged \citet{schafer2012scaling} and
 \citet{gauding2015line} to adopt Rice's theorem to estimate the
 number of extremal points $N_E$ of a turbulent signal~$\phi(x)$ as
 \begin{equation}
   \label{eq:rice2}
   N_E \propto \left[ - \frac{f''_{\phi_x}(0)}{f_{\phi_x}(0)}
   \right]^{1/2} \,,
 \end{equation}
 where $f_{\phi_x}$ is the normalized derivative correlation function
 defined as
 \begin{equation}
   f_{\phi_x}(r) = \frac{\avg{\phi_x(x) \phi_x(x+r)}}{\avg{\phi_x^2}} \,.
 \end{equation}
 Equation~\eqref{eq:rice2} makes use of the fact that the extremal
 points of $\phi(x)$ turn into zero-crossings of the derivative signal
 $\phi_x(x)$. With eq.~\eqref{eq:rice2} at hand, we can deduce the
 scaling of the mean length $\ell_m$ solely from properties of the
 derivative correlation function $f_{\phi_x}$.

\begin{figure}
   \centering
   \includegraphics[width=0.66\linewidth]{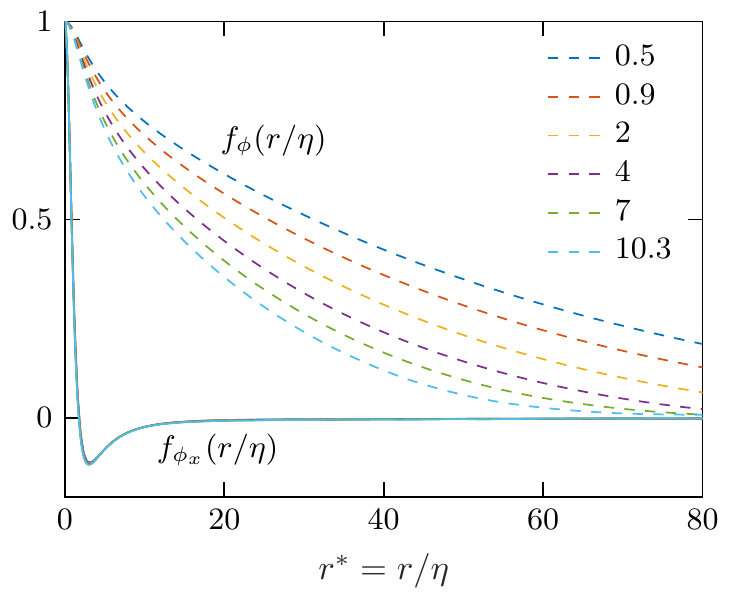}
   \caption{Normalized two-point correlation function of the scalar
     $f_\phi(r/\eta) = \avg{\phi(x+r) \phi(x)}/\avg{\phi^2}$ (dashed
     lines) and the scalar derivative
     $f_{\phi_x}(r/\eta) = \avg{\phi_x(x+r)
       \phi_x(x)}/\avg{\phi_x^2}$, displayed for the time steps
     indicated in table~\ref{tab:dns2}. The normalized scalar
     derivative two-point correlation function becomes self-similar
     when plotted as a function of $r^*=r/\eta$.}
   \label{fig:corr}
\end{figure}

 In fig.~\ref{fig:corr}, we display the normalized correlation
 function $f_\phi$ and the normalized derivative correlation function
 $f_{\phi_x}$ for different time steps during the self-similar decay
 as a function of $r/\eta$. Significant differences between
 $f_\phi(r/\eta)$ and $f_{\phi_x}(r/\eta)$ are evident, i.e.\ the
 correlation length for the scalar is much larger than that of the
 scalar gradient. The derivative correlation function
 $f_{\phi_x}(r/\eta)$ decays fast and becomes negative with a
 zero-crossing at $r/\eta \approx 1.6$ and a minimum at
 $r/\eta \approx 3.2$. This behavior can be explained from the general
 property that correlation functions of derivatives have zero integral
 length-scale $l_{\phi_x}$, i.e.
 \begin{equation}
   \label{eq:dcorr}
   l_{\phi_x} = \int_0^\infty f_{\phi_x}(r) {\rm d} r = - \int_0^\infty
   \dd{^2}{r^2} f_\phi(r) {\rm d} r
   = 0 \,,
 \end{equation}
 where the second equality follows from homogeneity of $\phi_x(x)$,
 and the third equality requires $f_\phi(r)$ to decay to zero for
 $r\to\infty$. A necessary condition following from
 eq.~\eqref{eq:dcorr} for $f_{\phi_x}(r/\eta)$ is the mutual
 cancellation between the areas below and above the abscissa. With
 these constraints, an almost perfect collapse of the normalized
 derivative correlation functions $f_{\phi_x}$ for the different time
 steps can be observed when plotted as a function of $r/\eta$, cf.\
 fig.~\ref{fig:corr}. The quality of the collapse is remarkable
 considering that during the reported period, the scalar gradient
 variance $\avg{\phi_x^2}$ and the Taylor-based Reynolds number
 $\mathit{Re}_\lambda$ decrease by a factor of 2.8 and 1.8,
 respectively.  The collapse of the normalized derivative correlation
 functions indicates complete self-similarity of
 $f_{\phi_x} (r/\eta)$. When $f_{\phi_x} (r/\eta)$ becomes
 self-similar by rescaling with a single length-scale $\eta$,
 eq.~\eqref{eq:rice2} requires with $\ell_m \propto 1/N_E$ a scaling
 of the mean length $\ell_m$ with the Kolmogorov length-scale $\eta$.

 Further justification for Kolmogorov scaling of the mean length
 $\ell_m$ can be provided by expressing the correlation functions in
 eq.~\eqref{eq:rice2} by one-point quantities, i.e.\
 $f_{\phi_x}(0) = \avg{\phi_x^2}$ and
 $f''_{\phi_{x}}(0) = - \avg{\phi_{xx}^2}$, and assuming KOC scaling
 for these quantities, 
 \begin{equation}
   \label{eq:phix}
 \avg{\phi_x^2}^{1/2} \propto D^{-1/2} \avg{\chi}^{1/2}
 \end{equation}
 and
 \begin{equation}
   \label{eq:phixx}
 \avg{\phi_{xx}^2}^{1/2} \propto \nu^{-1/4} D^{-1} \avg{\chi}^{1/2}
 \avg{\varepsilon}^{1/4} \,.
\end{equation}
Using \eqref{eq:phix} and \eqref{eq:phixx} in eq.~\eqref{eq:rice2},
confirms, with $N_E = 1/\ell_m$, Kolmogorov scaling for the mean
length $\ell_m$, i.e.
\begin{equation}
  \label{eq:ellm}
   \ell_m \propto \eta \propto \lambda \mathit{Re}_\lambda^{-1/2} \,,
 \end{equation}
 where we considered in \eqref{eq:ellm} a unity Schmidt number.

\begin{figure}
   \centering
   \includegraphics[width=0.66\linewidth]{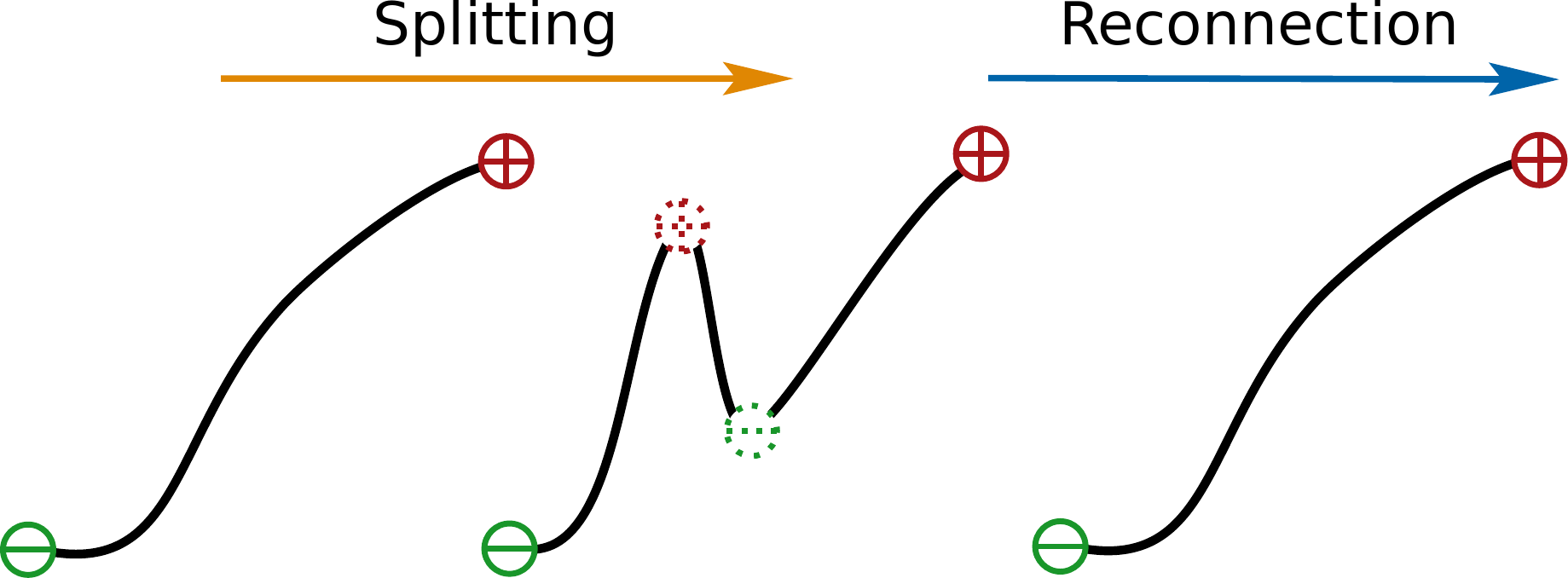}
   \caption{Illustration of a splitting and reconnection event. By
     splitting, a line segment is divided into three smaller segments
     by introducing new extremal points. Reconnection refers to an
     event when extremal points disappear, i.e.\ due to diffusion, and
     smaller elements are joint to form one large segment. }
   \label{fig:fast}
\end{figure}

The temporal evolution of line segments in decaying turbulence is
governed by a complex process, involving slow and fast changes as
discussed by \citet{wang2006length} and \citet{schaefer2009fast}. Slow
changes are responsible for a continuous evolution of the ending
points. As the ending points are moved relatively to each other, slow
changes can modify the length distribution function $P(\ell)$ but not
the mean length $\ell_m$. On the other hand, fast changes describe
processes where the connectivity of line segments changes abruptly due
to a topology change of the scalar field. An abrupt topology change
occurs when local extremal points disappear due to diffusion or when
new local extremal points appear due to the stretching and folding
mechanism of turbulent flows. This process results in a discrete
change of both $\Delta \phi$ and $\ell$, and can hence modify the mean
length $\ell_m$. \citet{wang2006length} distinguished between two
different fast processes.  By a reconnection process, two initially
independent dissipation elements are merged and a new dissipation
element with a larger $\Delta \phi$ is created.  A splitting process,
on the other hand, generates new local extremal points resulting in a
shortening of line segments and in an abrupt decrease of
$\Delta \phi$. Figure \ref{fig:fast} illustrates a cutting and
reconnection process. Following \citet{wang2008phd}, the mean length
of line segments is governed by the evolution equation
 \begin{equation}
   \frac{1}{\ell_m} \dd{\ell_m}{t} = \ell_m \Lambda_1 - 2 \Lambda_2
 \end{equation}
 where $\Lambda_1$ and $\Lambda_2$ denote the frequencies of
 generation and annihilation of extremal points,
 respectively. Physically, $\Lambda_1$ is determined by the turbulence
 intensity, while $\Lambda_2$ is determined by the molecular
 diffusivity and the mean length-scale.  In decaying turbulence, the
 turbulence intensity decreases and molecular diffusion smooths the
 scalar field.  As a consequence, the annihilation process prevails
 over the generation process leading to the observed increase of the
 mean length $\ell_m$, or, the decrease of the number of extremal
 points~$N_E$.

\section{The joint distribution of $\Delta \phi$ and $\ell$}
\label{sec:joint}
Turbulent flows are characterized by a complex spatio-temporal
structure with a non-local interaction between various
length-scales. Conventionally, more understanding of the spatial flow
structure can be obtained from two-point or multi-point statistics,
rather than one-point statistics.  Line segments are parameterized by
the distance $\ell$ and the scalar difference $\Delta \phi$ between
adjacent extremal points. Then, most statistical properties are
captured by the joint statistics of these parameters. The Bayes
theorem relates the marginal pdf $P(\ell)$ of the length of line
segments to the joint and conditional pdfs by
\begin{equation}
  P_c (\Delta \phi | \ell) = \frac{P(\Delta\phi,\ell)}{P(\ell)}\,,
\end{equation}
where $P_c(\Delta\phi | \ell)$ refers to the conditional pdf, and
$P(\Delta\phi,\ell)$ is the joint probability density function (jpdf).

\begin{figure}
   \centering
   \includegraphics[width=0.66\linewidth]{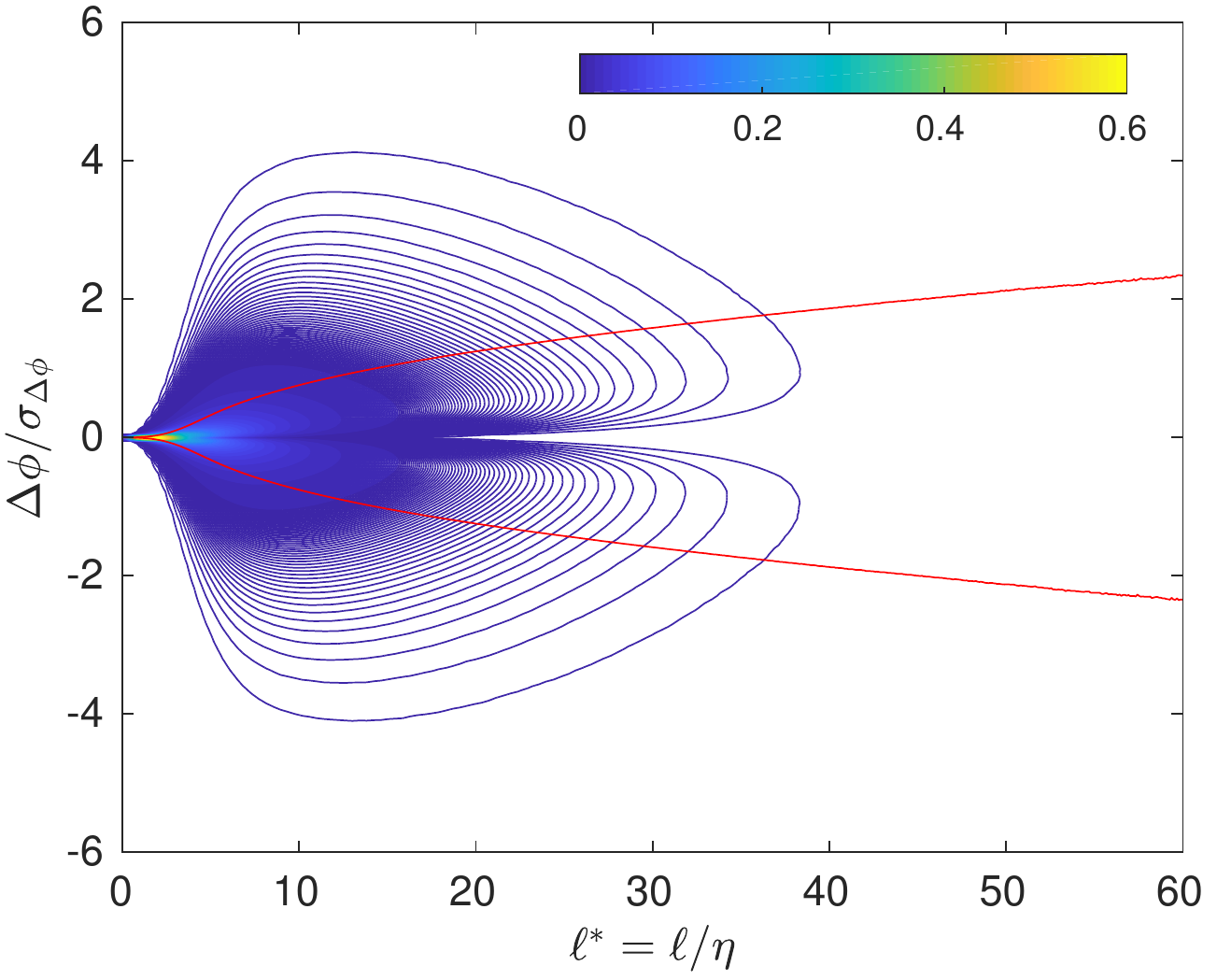} \\[1ex]
   \includegraphics[width=0.66\linewidth]{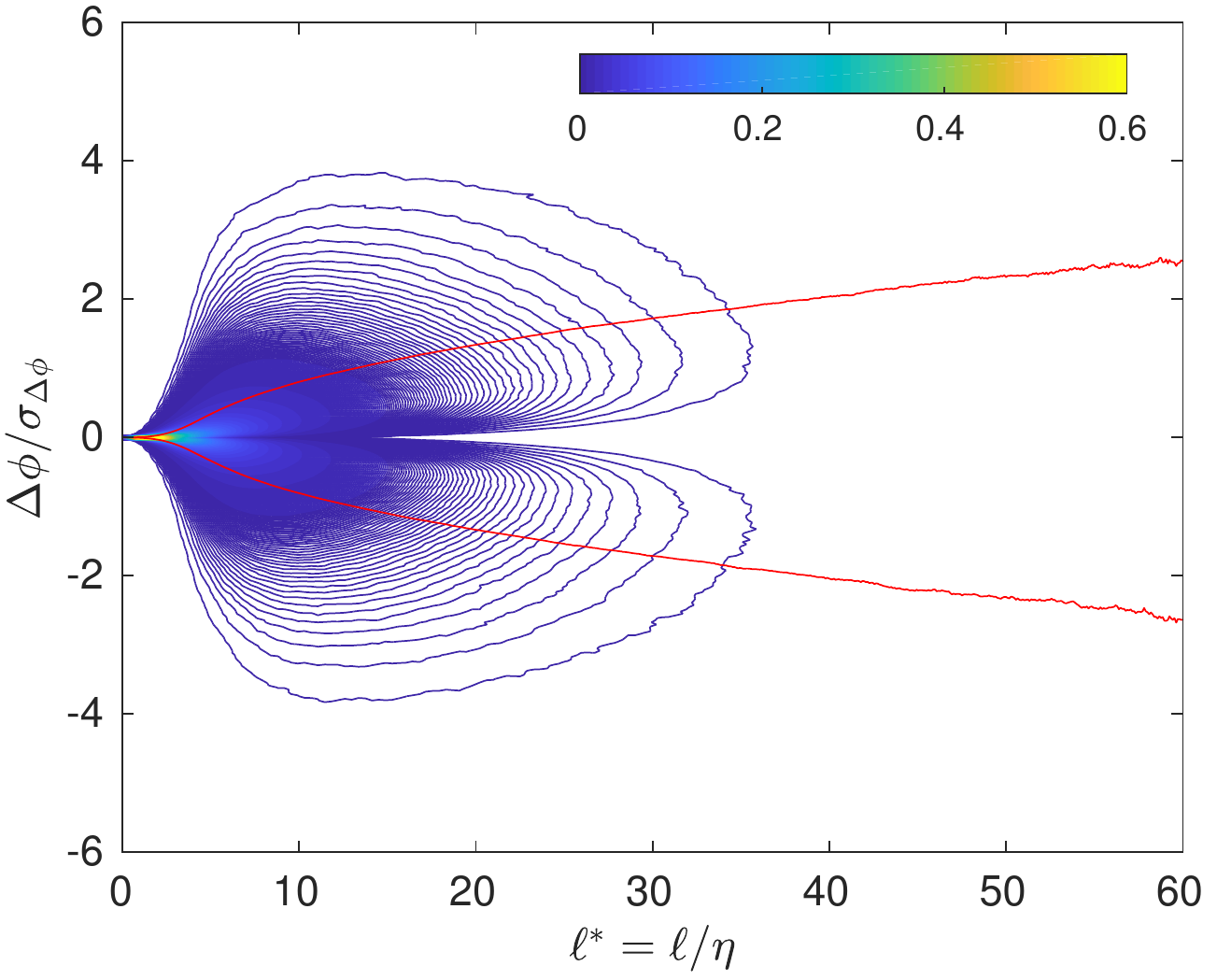} 
   \caption{Joint distribution function
     $P(\Delta\phi,\ell)$ for time step D2 (top) and time step D6. The
     red lines refer to the normalized conditional average
     $\avg{\Delta\phi|\ell}$,  which are shown separately
     for  positive and negative wings.  }
   \label{fig:jpdf}
 \end{figure}

Figure~\ref{fig:jpdf} shows the normalized joint probability density
function $P(\Delta\phi,\ell)$ for two different time steps D1 and D6
during the self-similar decay. The abscissa is normalized by the
Kolmogorov length $\eta$ and the ordinate is normalized by the
standard deviation $\sigma_{\Delta\phi}$, given by
$\sigma_{\Delta\phi} = \avg{(\Delta\phi)^2}^{1/2}$. The jpdf exhibits
two distinct wings, where the upper wing corresponds to positive
segments with $\Delta\phi>0$, and the lower wing corresponds to
negative segments with $\Delta\phi<0$.  The scale-dependent
  skewness $\avg{(\Delta\phi)^3|\ell}/\avg{(\Delta\phi)^2|\ell}^{3/2}$
  is close to zero for all scales $\ell$, which signifies that the
  wings are symmetric with respect to the abscissa. 
Figure~\ref{fig:jpdf} also displays the conditional normalized mean
scalar difference $\avg{\Delta\phi| \ell/\eta}/\sigma_{\Delta\phi}$
separately for positive and negative segments. The magnitude of the
conditional mean increases monotonously with $\ell/\eta$.  It can be
concluded that $\Delta\phi$ and $\ell$ are not independent and that on
average, large segments also have a large scalar difference. The jpdf
$P(\Delta\phi,\ell)$ covers different physical effects. The upper and
lower left corners, where $\ell$ is small but the magnitude of
$\Delta\phi$ is large, represent poorly mixed regions, characterized
by large absolute values of the mean gradients
$g=\Delta\phi/\ell$. Well mixed regions exist close to the abscissa,
where the magnitude of $\Delta\phi$ is small and $\ell$ stays
sufficiently large. Comparing the normalized jpdfs for time steps D2
and D5 reveals that the shape of the jpdfs is very similar but not
completely self-similar. At the late time step, segments with very
large length-scale $\ell/\eta$ and segments with very large scalar
difference $| \Delta \phi |/\sigma_{\Delta\phi}$ are slightly less
significant. This can be explained by the decrease of the Reynolds
number and the associated smoothing of the turbulent field. The
consequence of this observation on the self-similarity of line
segments will be explored in more detail in the next section.

\begin{figure}
   \centering
   \includegraphics[width=0.66\linewidth]{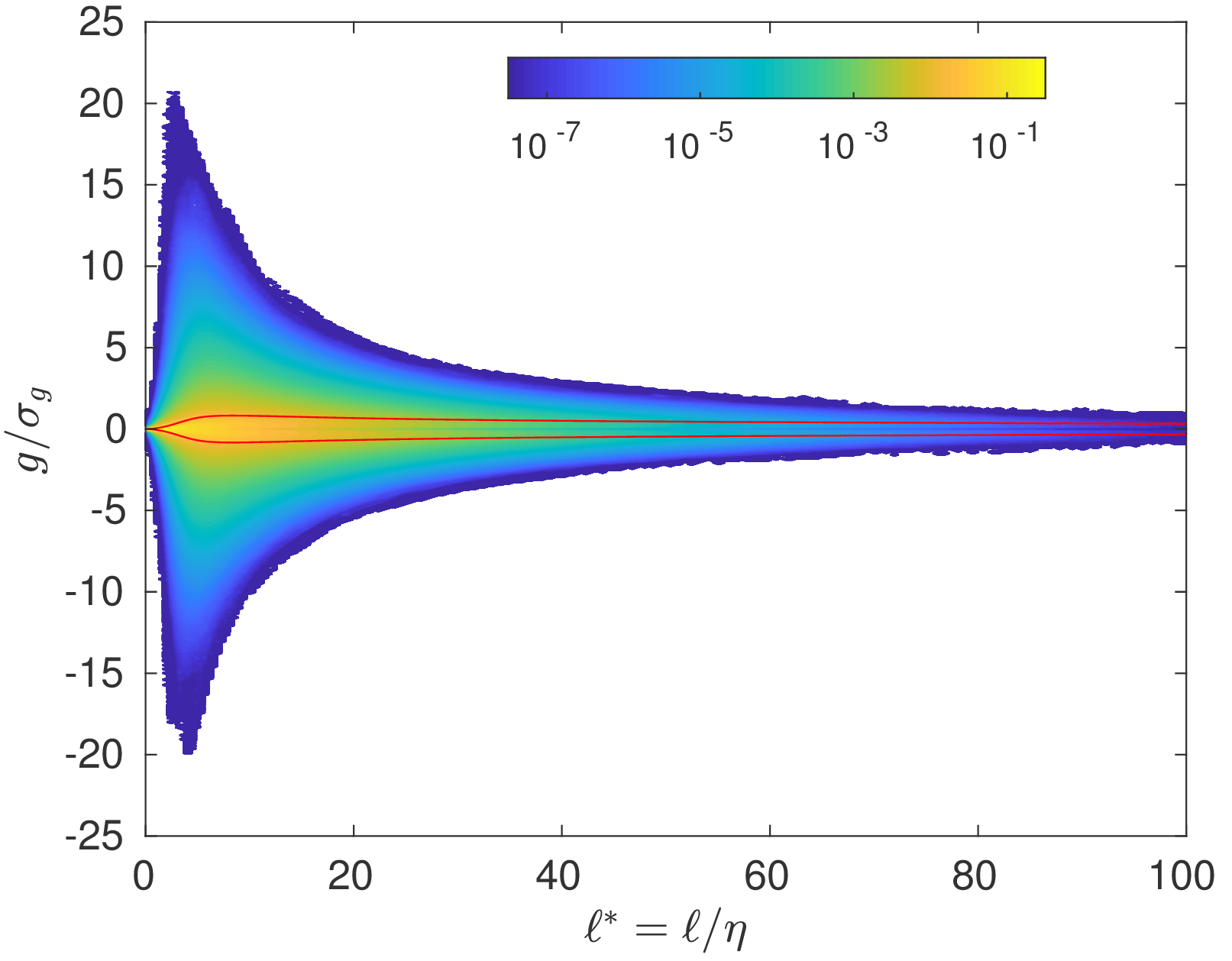} \\[1ex]
   \includegraphics[width=0.66\linewidth]{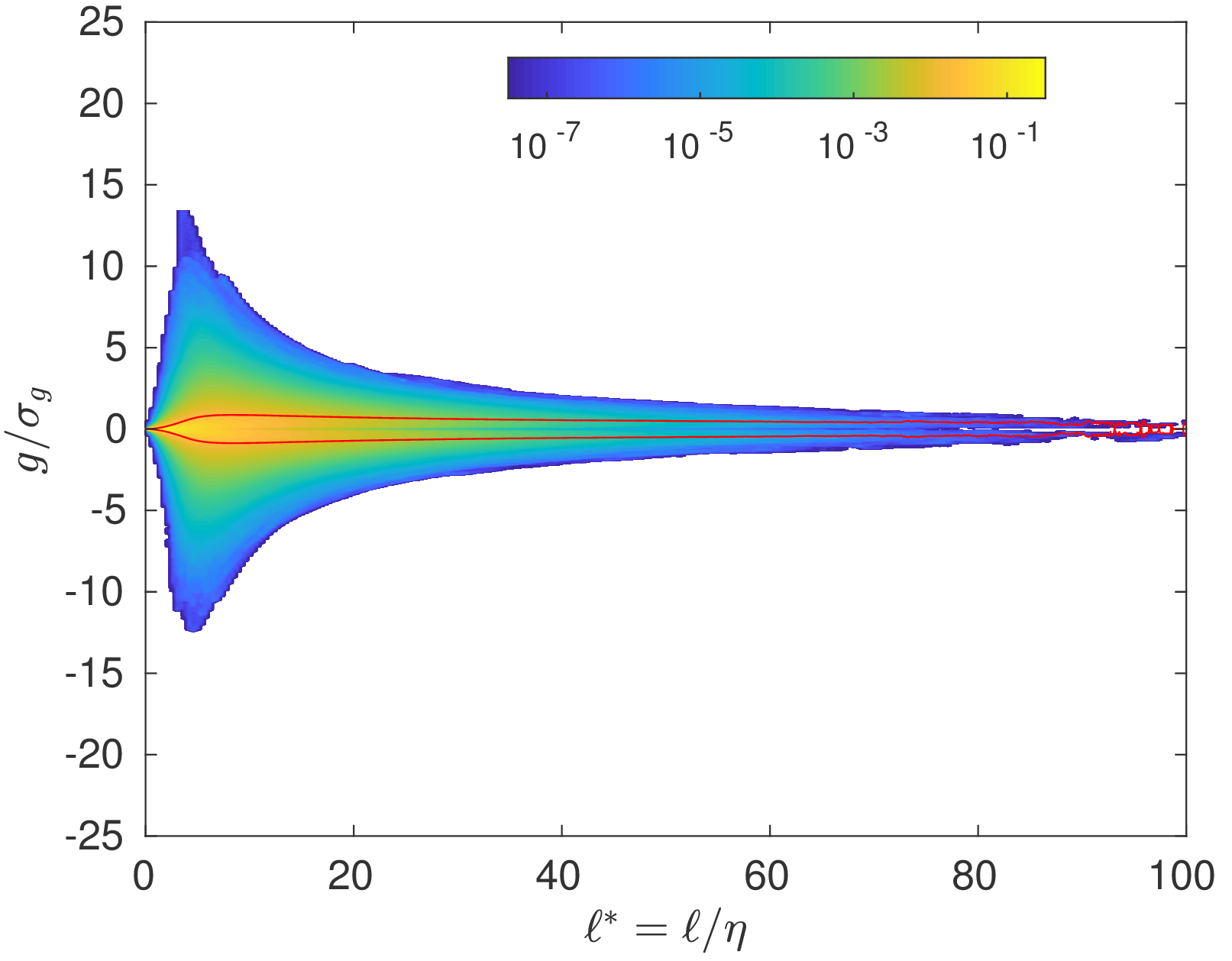} 
   \caption{Joint distribution function $P(g,\ell)$ for time step D2
     (top) and time step D6. The red lines refer to the normalized
     conditional average $\avg{g|\ell}$, which are shown separately
     for the positive and negative wings. }
   \label{fig:jpdf_g}
 \end{figure}

 To study turbulent mixing, a line segment based mean gradient can be
 defined as
\begin{equation}
  g = \frac{\Delta \phi}{\ell} \,,
\end{equation}
which is proportional to the mean diffusive flux or the mean local
scalar gradient between adjacent extremal points, i.e.
\begin{equation}
  \label{eq:g}
  \frac{1}{\ell} \int_{x_1}^{x_1+\ell} \left( D \pp{\phi}{x_1} \right)
  {\rm d} x_1 =
  D\frac{\Delta\phi}{\ell} \,.
\end{equation}
Dimensionally, the mean gradient $g$ can be related to the scalar mean
dissipation, i.e.
\begin{equation}
  \avg{g^2} \propto \frac{\avg{\chi}}{D} \,.
\end{equation}
With these properties, the jpdf $P(g,\ell)$ provides information about
the scale-dependence of turbulent mixing. The normalized jpdf
$P(g,\ell)$ is displayed for two different time steps D2 and D6 in
fig.~\ref{fig:jpdf_g}.  The abscissa is normalized by the Kolmogorov
length $\eta$ and the ordinate is normalized by the rms of $g$,
defined as $\sigma_g = \avg{g^2}^{1/2}$. Similar to
$P(\Delta\phi,\ell)$, the jpdf $P(g,\ell)$ has two distinct wings for
positive and negative gradients. The wings are symmetric and reveal
with respect to the abscissa a long tail which is situated at small,
but finite length-scale $\ell$. That means that large gradients occur
at a finite length-scale and not in the limit $\ell\to 0$.  This
behavior comes from the fact that $\Delta \phi$ and $\ell$ are not
independent.  At small scales, $\Delta\phi$ tends rapidly to zero due
to molecular damping resulting in vanishing values of $g$.  It is
important to emphasize that the scale-dependent scalar gradient based
on the scalar increment $\delta \phi$ has a non-zero limit for
$r\to 0$, as the compensated structure function
\begin{equation}
  \frac{\avg{(\delta \phi)^{2n}}}{r^{2n}}
\end{equation}
turns into the corresponding moment of local scalar gradient,
\begin{equation}
  \lim_{r\to 0} \frac{\avg{(\delta \phi)^{2n}}}{r^{2n}} =
  \avg{\left(\pp{\phi}{x_1} \right)^{2n}} \,,
\end{equation}
in the small-scale limit for $r \to 0$. 

 Figure \ref{fig:jpdf_g} reveals that long line segments are
   predominantly characterized by small mean gradients $g$. These line
   segments represent relatively well mixed regions, which appear as
   ramp-like structures in the scalar field. For the present DNS, the
   size of these coherent regions reaches up to $100 \eta$.

   On the other hand, very large positive and negative values of $g$
   can be found at the tips of the wings, which exceed the rms value
 of $g$ by a factor close to 20 (for case D2) and close to 15 (for
 case D6). This indicates the existence of strong internal
 intermittency. In scalar fields, intermittency results from the
 characteristic cliff-like structure \cite{shraiman2000scalar}, formed
 by a straining motion, which has its origin in the vortex-stretching
 mechanism of turbulence. As observed in fig.~\ref{fig:jpdf_g}, the
 normalized jpdf $P(g,\ell)$ is clearly not self-similar, because
 intermittency weakens during the decay. This result confirms the
 standard paradigm of turbulence that especially the rare large events
 break self-similarity.

\begin{figure}
   \centering
   \includegraphics[width=0.66\linewidth]{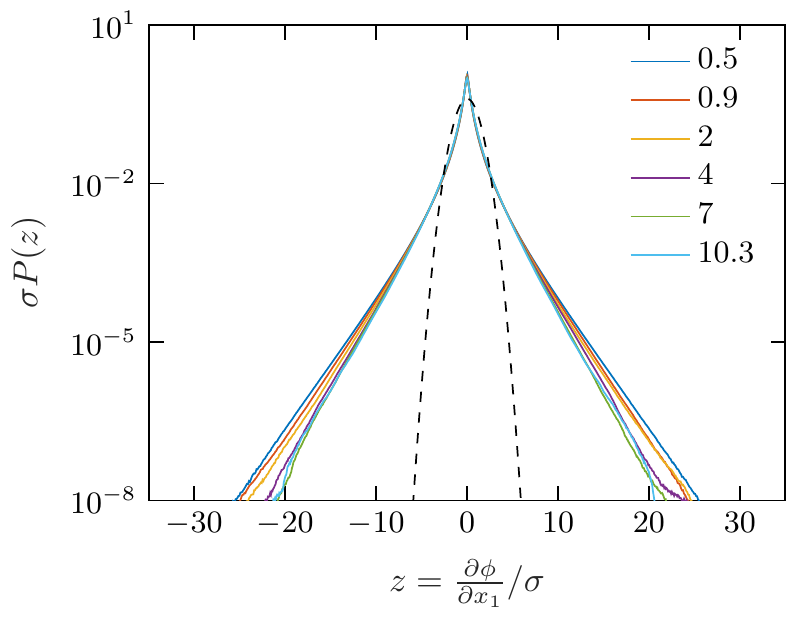}
   \caption{Marginal pdf of the scalar gradient
     $\partial \phi / \partial x_1$ for different time steps
     normalized by the standard deviation
     $\sigma={\avg{(\partial \phi/ \partial x_1)^2}}^{1/2}$ for each
     curve.  The black dashed curve represents a standard normal
     distribution.}
   \label{fig:pdf_phix}
 \end{figure}

\begin{figure}
   \centering
   \includegraphics[width=0.66\linewidth]{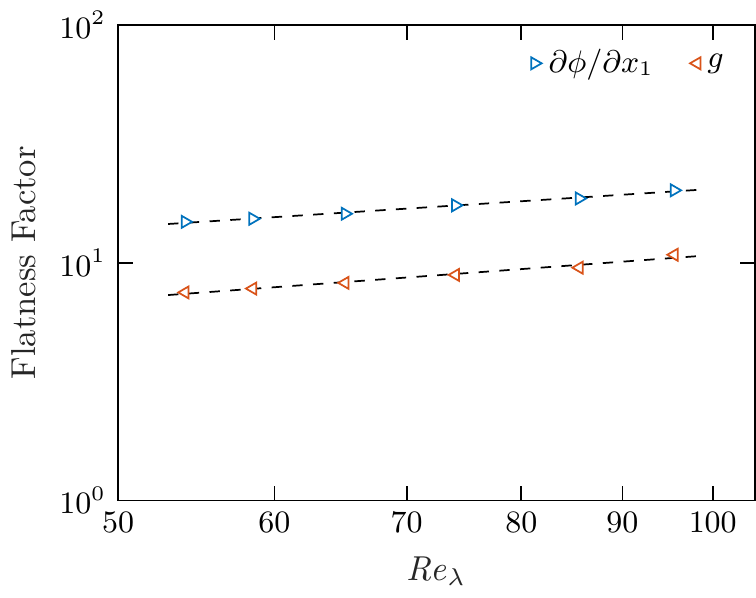}
   \caption{Scaling of the flatness factor of the local scalar
     gradient $\partial \phi / \partial x_1$ and the segment based
     mean gradient $g=\Delta\phi/\ell$ as a function of
     $\mathit{Re}_\lambda$. The black dashed lines indicate a
     least-square fit shown for reference.}
   \label{fig:flat}
 \end{figure}

Intermittency effects are also observed from the pdf of the local
scalar gradient $P(\partial\phi/x_1)$, displayed in
fig.~\ref{fig:pdf_phix}.  The normalized pdfs of the scalar gradient
are strongly non-gaussian and exhibit stretched exponential tails. The
tails of the pdfs, which represent rare extreme events, do not
collapse for different time steps: they become less pronounced during
the decay and are clearly not self-similar. To quantify the
statistical behavior of rare extreme events, the local gradient
flatness
\begin{equation}
  F=\frac{\avg{\left( \pp{\phi}{x_1} \right)^4}}{\avg{\left(
        \pp{\phi}{x_1} \right)^2}^2} \,,
\end{equation}
and the flatness of the mean gradient of line segments $g$
\begin{equation}
  F_g = \frac{\avg{g^4}}{\avg{g^2}^2} \,, 
\end{equation}
can be introduced as the normalized fourth-order moment
\cite{frisch1995}. For line segments, the even order moments of the
mean gradient $g$ are given by
\begin{equation}
  \avg{g^{2n}} = \int_{-\infty}^{\infty} \int_0^\infty g^{2n} P(g,\ell) {\rm
    d} \ell {\rm d}g \,.
\end{equation}
Figure \ref{fig:flat} illustrates the scaling of $F$ and $F_g$ as a
function of the Reynolds number $\mathit{Re}_\lambda$. From the DNS
results, it follows that $F$ and $F_g$ can be approximated by a
power-law, i.e.\ $\propto \mathit{Re}_\lambda^\alpha$ where the
scaling exponent $\alpha$ is close to 0.55 for both
quantities. Despite virtually the same scaling exponent, $F_g$ is
considerably smaller than $F$, as the segment based gradient $g$ is
already an average quantity defined over a stochastic length-scale
$\ell$. The flatness for both quantities increase with Reynolds number
(or alternatively decrease with time). A constant flatness factor is a
necessary condition for complete self-similarity, so that we can
conclude that in this kind of flow, neither the local gradient
$\partial \phi /\partial x_1$ nor the mean gradient of line segments
$g$ are self-similar. Note that under the conditions of the KOC
theory, which hypothesizes universality of small-scale turbulence, a
constant flatness factor is predicted
\cite{antonia2015boundedness,tang2018reappraisal}.

\begin{figure}
   \centering
   \includegraphics[width=0.66\linewidth]{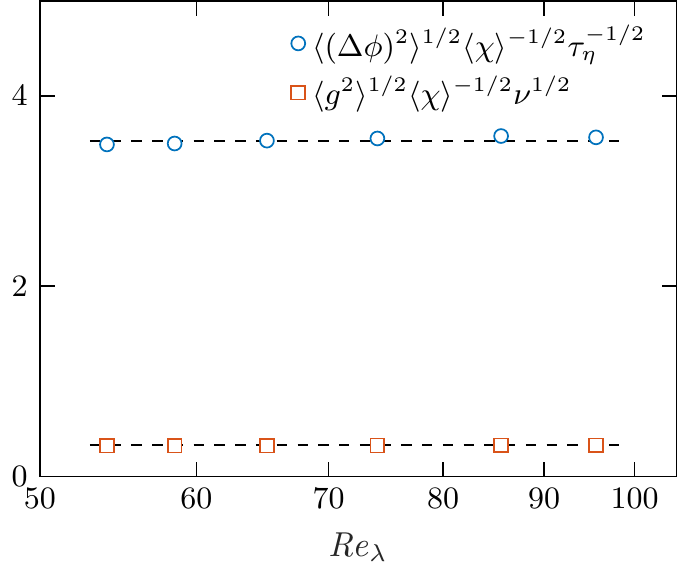} 
   \caption{Reynolds number dependence of the normalization quantities
     $\sigma_{\Delta\phi} = \avg{(\Delta\phi)^2}^{1/2}$ and
     $\sigma_{g} = \avg{g^2}^{1/2}$. The dashed lines represent
     $\sigma_{\Delta\phi} \approx 3.5$ and $\sigma_g\approx 0.3$,
     respectively. }
   \label{fig:scaling_sigma}
 \end{figure}

 For reference, fig. \ref{fig:scaling_sigma} shows the Reynolds number
 dependence of the normalization quantities $\sigma_{\Delta\phi}$ and
 $\sigma_g$. For the investigated range of Reynolds numbers, both
 quantities scale with good accuracy with the KOC quantities, i.e.\
 $  \sigma_{\Delta\phi} \propto \avg{\chi}^{1/2} \tau_\eta^{1/2} $
 and
 $ \sigma_g \propto \avg{\chi}^{1/2} \nu^{-1/2} $,
 with the Kolmogorov time-scale given by
 $\tau_\eta=(\nu/\avg{\varepsilon})^{1/2}$.
 
\section{Scaling of the conditional averages $\avg{\Delta\phi|\ell}$
  and $\avg{\Delta u |\ell}$}
\label{sec:cond}
An $n$th order conditional average of line segments,
\begin{equation}
  \label{eq:sf}
  \avg{|\Delta\phi|^n|\ell}  = \int_{-\infty}^\infty |\Delta\phi|^n
  \frac{P(\Delta\phi,\ell)}{P(\ell)} {\rm d} (\Delta\phi) \,,
\end{equation}
can be defined. In contrast to the definition of conventional
structure functions, i.e.\
\begin{equation}
  \avg{| \delta \phi |^n }  = \avg{| \phi(x+r) - \phi(x) |^n } \,,
\end{equation}
the distance $\ell$ in \eqref{eq:sf} is not arbitrarily chosen, but
rather determined by the local structure of turbulence, represented by
the distance $\ell$ and the fluctuations $\Delta\phi$ between adjacent
local extremal points. The conditional mean
$\avg{|\Delta\phi|^n|\ell}$ is computed for segments belonging to the
same length-class $\ell$, rather than for arbitrary points with the
same separation distance $r$.

Under the assumptions of the KOC theory, statistics of $\Delta\phi$
are uniquely determined by the mean energy dissipation
$\avg{\varepsilon}$, the mean scalar dissipation $\avg{\chi}$ and the
molecular diffusivity $D$. As a straightforward consequence, the
conditional average $\avg{|\Delta \phi|^n | \ell}$ can be expressed by
functional forms,
\begin{equation}
  \avg{|\Delta \phi|^n | \ell^*} = A_n(t) f_n(\ell^*) \,,
\end{equation}
that are built as products between a time and order-dependent
prefactor $A_n(t)$ and a order-dependent shape function $f_n(\ell^*)$
of the normalized distance $\ell^*$. Following the previous findings,
we define $\ell^* = \ell/\eta$, with $\eta \propto \ell_m$ being the
characteristic length-scale.

\begin{figure}
   \centering
   \includegraphics[width=0.66\linewidth]{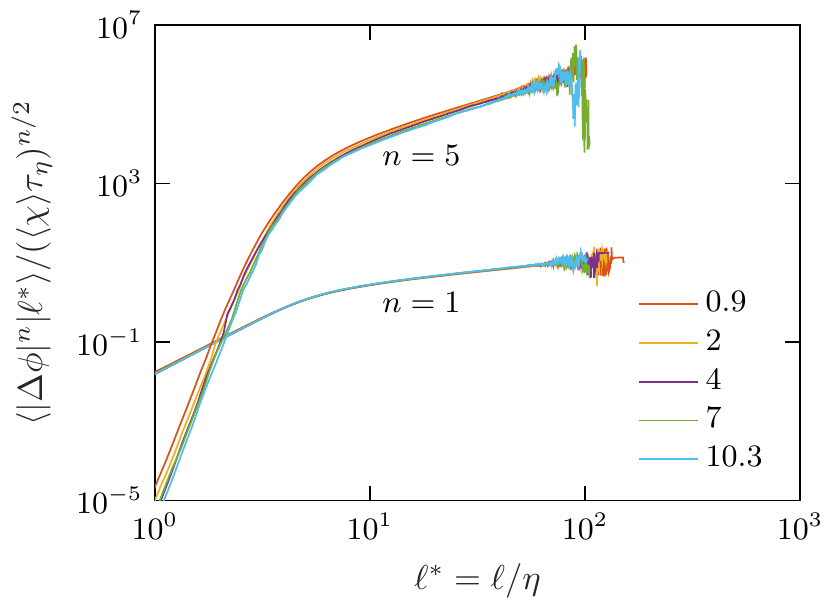} 
   \caption{Normalized conditional scalar difference
     $ \avg{\Delta \phi |^n \ell} $ for different time steps and for
     $n=1$ and $n=5$.   }
   \label{fig:cmean}
 \end{figure}

Figure \ref{fig:cmean} shows the normalized conditional mean
$ \avg{|\Delta \phi|^n | \ell} $ for different time steps for $n=1$
and $n=5$. Similar to structure functions, the conditional mean
exhibits two distinct scaling regimes for small scales in the
dissipative range and for larger scales in the inertial
range. Normalized with KOC quantities, i.e.\
$(\avg{\chi} \tau_\eta)^{n/2}$, the curves collapse, indicating
self-similarity. However, the collapse is much better for the first
order $n=1$ than for the fifth order $n=5$. This finding is in
agreement with the observation that the normalized jpdf of
$\Delta\phi$ and $\ell$ is not completely self-similar for very large
values of $|\Delta\phi|$ or $\ell$. With eq.~\eqref{eq:sf}, these
parts of the jpdf mostly contribute to higher-order statistics
affecting the quality of collapse for the fifth-order structure
function. For the first-order conditional average
$ \avg{|\Delta \phi| | \ell} $, primarily the inner parts of
$P(\Delta\phi,\ell)$, which are self-similar, contribute to the
average.

\begin{figure}
   \centering 
   \includegraphics[width=0.66\linewidth]{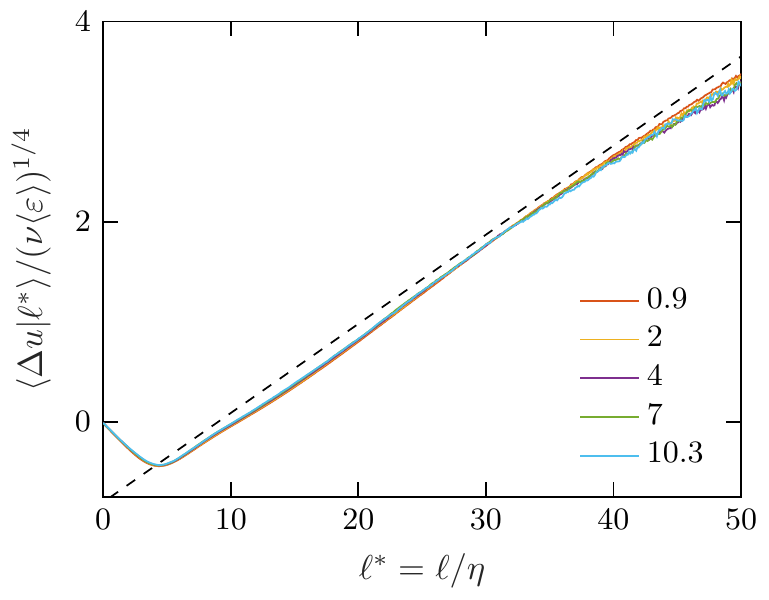} 
   \caption{Normalized conditional velocity difference
     $ \avg{\Delta u| \ell} $ for different time steps. }
   \label{fig:cmean_du}
 \end{figure}

 One of the most important result derived from the incompressible
 Navier-Stokes equations is the Kolmogorov equation for the velocity
 structure function \cite{kolmogorov1941}. Under the
 condition of local isotropy and for sufficiently high Reynolds
 numbers, Kolmogorov hypothesized
\begin{equation}
  \label{eq:k41}
\avg{|{\delta u}|} \propto (\avg{\varepsilon} r )^{1/3} \,,
\end{equation}
where $\delta u$ is the velocity increment taken in longitudinal
direction. From a different viewpoint, the velocity difference
$\Delta u$ can be defined with respect to adjacent local extremal
points $x_{\rm start}$ and $x_{\rm end}$ of the scalar field
\cite{wang2009scaling}, i.e.
\begin{equation}
  \avg{\Delta u | \ell} = \avg{u(x_{\rm end}) - u(x_{\rm
      start})|x_{\rm end}-x_{\rm start}=\ell} \,.
\end{equation}
The velocity difference $\Delta u$ is directly connected to a
straining motion that is acting on the scalar field. By a straining
motion, scalar gradients are smoothed in the direction of extensive
strain and steepened in the direction of compressive strain. The
balance between strain and diffusion generates the characteristic
cliff-ramp-like structure observed in scalar fields.
Figure~\ref{fig:cmean_du} displays the velocity difference
$\avg{\Delta u| \ell}$ for the different time steps. It can be
observed that small segments are subject compressive strain
($\Delta u<0$), while larger segments ($\ell/\eta > 10$) are subject
to extensive strain ($\Delta u >0$). By normalization with the
Kolmogorov velocity $u_\eta=(\nu\avg{\varepsilon})^{1/4}$ and the
Kolmogorov length-scale $\eta$, the curves reveal an excellent
collapse for the different time steps indicating complete
self-similarity. The increase of $\avg{\Delta u| \ell}$ at larger
scales, i.e.\ $\ell/\eta>10$, is linear, and fig.~\ref{fig:cmean_du}
suggests the scaling relation
\begin{equation}
  \label{eq:du}
  \avg{\Delta u | \ell } \propto \frac{1}{\tau_\eta} \ell \,.
\end{equation}
Compared with Kolmogorov's conventional $r^{1/3}$ scaling,
cf.~\eqref{eq:k41}, a linear scaling with $\ell/\tau_\eta$ is
surprising. It can be explained from the fact that large coherent
regions with monotonously varying scalar values are formed
preferentially by extensive strain. From dimensional grounds, the
straining motion in turbulence scales with the Kolmogorov time
$\tau_\eta$, which leads to eq.~\eqref{eq:du}.


 \section{Conclusion}
 \label{sec:conclusion}
 The self-similarity of scalar mixing in decaying homogeneous
 isotropic turbulence has been studied by the method of line
 segments. By decomposing in one-dimensional space, the method of line
 segments decomposes the scalar field into smaller sub-units based on
 local extremal points. Within these sub-units, the value of the
 scalar varies monotonously. The segments were parameterized by the
 length $\ell$ and the scalar difference $\Delta\phi$ between the
 ending points.  Line segments can be understood as thin
   local diffusive-convective structures, whose average length
   $\ell_m$ equals approximately 10 times the Kolmogorov length scale
   $\eta$.  The analysis was based on a highly resolved direct
 numerical simulation.  The main findings are:
 \begin{enumerate}
 \item The length distribution of line segments $P(\ell)$ is strongly
   non-gaussian. It exhibits a strong dependence on time reflecting
   the increase of the mean length $\ell_m$ during the decay. After
   normalization with the mean length the rescaled pdf
   $\tilde P(\tilde \ell)$ becomes completely self-similar. A
   statistical analysis signifies that the mean length $\ell_m$ scales
   with the Kolmogorov length-scale $\eta$.
 \item Further information about the local structure of turbulence is
   provided by joint statistics of the scalar difference $\Delta\phi$
   and the length $\ell$. Both quantities are correlated and long
   segments have on average a large scalar difference
   $|\Delta\phi|$. The normalized joint pdf reveals a self-similar
   core, but regions of very large $\ell$ or very large $|\Delta\phi|$
   become slightly less significant during the decay.
 \item To investigate turbulent mixing, a line segment based gradient
   was defined as $g=\Delta\phi/\ell$.  Large values of $g$ occur at a
   small, but finite length-scale. Due to internal intermittency, the
   joint pdf of $g$ and $\ell$ exhibits a long tail representing large
   segment based gradients. These intense gradients stem from
   cliff-like structures, which are formed by a straining motion. The
   joint pdf of $g$ and $\ell$ is clearly not self-similar as
   gradients with large magnitude disappear during the decay.
 \item Consistent with the jpdf $P(\Delta\phi,\ell)$, the normalized
   conditional average of $\avg{\Delta\phi^n|\ell}$ becomes completely
   self-similar for low orders. Complete self-similarity was not
   observed for higher-orders, which is in agreement with the standard
   paradigm of turbulence that higher-order statistics reflect the
   effect of internal intermittency. The conditional average of the
   velocity difference, i.e.\ $\avg{\Delta u |\ell}$, allows to assess
   the impact of a straining motion on the scalar field. In agreement
   with existing theories, we found that small scales are subject to a
   compressive strain, while larger length-scales, over which the
   scalar values varies monotonously, are subject to an extensive
   strain. It was shown that the normalized conditional velocity
   difference is self-similar and obeys a linear scaling with the
   Kolmogorov length and time scales.
 \end{enumerate}

\appendix

\section{Anisotropy of the scalar field}
\label{sec:aniso}  
\begin{figure}
   \centering
   \includegraphics[width=0.66\linewidth]{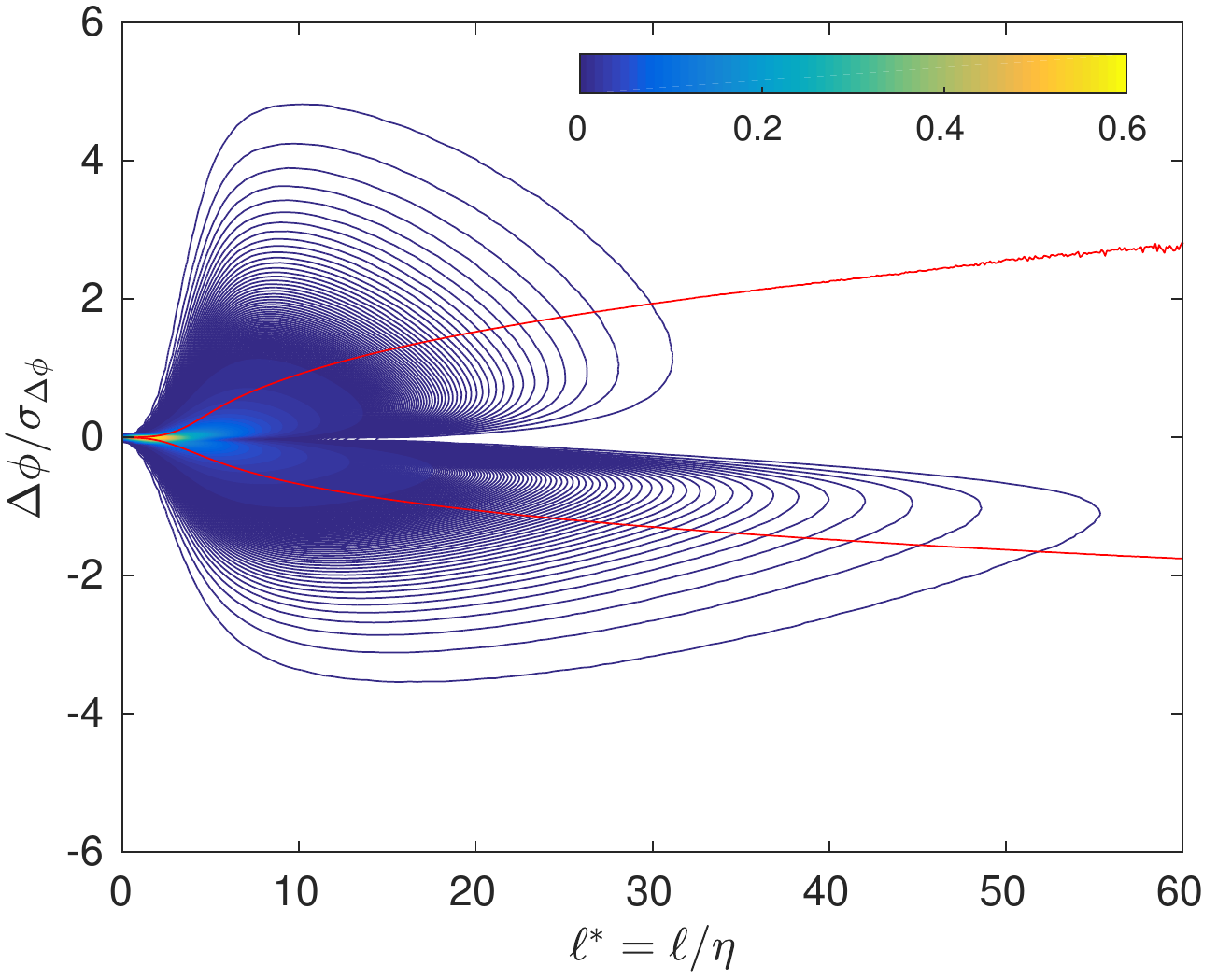} \\[1ex]
   \includegraphics[width=0.66\linewidth]{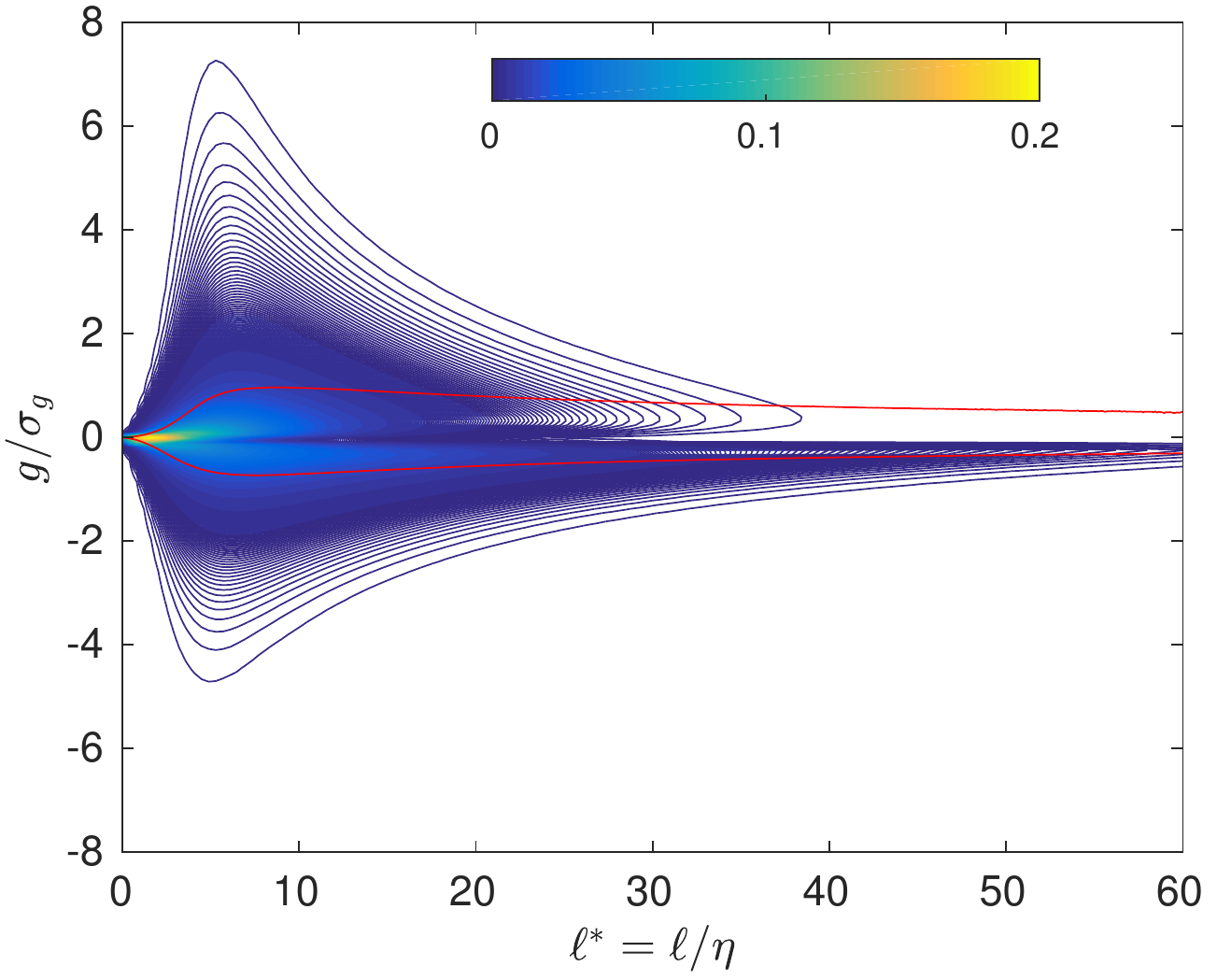} 
   \caption{ Joint distribution function $P(\Delta\phi,\ell)$
     (top) and joint distribution function $P(g,\ell)$ (bottom) for
     time step D2. Line segments are computed in $x_2$-direction
     parallel to the mean scalar gradient. The red lines refer to the
     normalized conditional averages $\avg{\Delta\phi|\ell}$ and
     $\avg{g|\ell}$, respectively, which are shown separately for
     positive and negative wings. }
   \label{fig:jpdf_y}
 \end{figure}

As the scalar field is not isotropic,
  directional statistics are of interest and provide a deeper insight
  into cliff-ramp-like structures. Figure~\ref{fig:jpdf_y} displays
  the normalized joint pdfs $P(\Delta\phi,\ell)$ and $P(g,\ell)$
  computed for line segments in $x_2$-direction (parallel to the mean
  scalar gradient).  The jpdfs display a clear asymmetry with respect
  to the abscissa, which is different from line segments in
  $x_1$-direction (perpendicular the the mean scalar gradient),
  cf.~figs.~\ref{fig:jpdf} and \ref{fig:jpdf_g}.
  Figure~\ref{fig:jpdf_y} reveals that increasing line segments
  ($\Delta\phi>0$) are on average shorter and have a larger scalar
  difference compared to decreasing line segments.  On the other hand,
  decreasing line segments ($\Delta\phi<0$) are on average longer and
  have a smaller absolute scalar difference. As a consequence, it is more
  likely that increasing  line segments appear as cliff-like structures,
  while decreasing line segments appear as ramp-like structures. This
  finding reflects the characteristic cliff-ramp-like structures that exist in the
  direction parallel to the mean scalar gradient and is in agreement
  with the positive scalar gradient skewness $S$ displayed in
  fig.~\ref{fig:skew}.

\section*{Acknowledgment}

Financial support was provided by the Labex EMC3, under the grant
VAVIDEN, as well as the Normandy Region and FEDER. Additionally, the
authors gratefully acknowledge the computing time granted on the
supercomputer JUQUEEN (Research Center Juelich
\cite{stephan2015juqueen}). The authors would like to thank Dr.\
Michael Stephan from Juelich Supercomputing Center for his continuous
support that helped us to perform simulations with more than 1.8
Million threads.


\section*{References}
\bibliographystyle{elsarticle-harv} 
\bibliography{main}

\end{document}